\RequirePackage{lineno}
\documentclass[prb,twocolumn,showpacs,a4paper]{revtex4}
\usepackage{graphicx}
\usepackage{dcolumn}
\usepackage{natbib}
\usepackage[english]{babel}
\usepackage{lineno}
\usepackage{epsfig}
\usepackage{appendix}
\usepackage{multirow}

\begin{document}
\newcommand{\putgraph}[2]{\includegraphics[#1,keepaspectratio]{#2.pdf}} 
\newcommand{\putif}[2]{\includegraphics[#1,keepaspectratio]{#2}}
\newcommand{\putiff}[2]{\parbox[c]{#1}{\includegraphics[height=#1,keepaspectratio]{#2}}}

\newcommand{\putifn}[1]{\parbox[c]{2cm}{\includegraphics[height=2cm,keepaspectratio]{#1}}}
\newcommand{\putifm}[1]{\parbox[c]{3cm}{\includegraphics[height=3cm,keepaspectratio]{#1}}}

\newcommand{\fig}[1]{Fig.\ref{fig:#1}}
\newcommand{\figa}[2]{Fig.\ref{fig:#1}(#2)}
\newcommand{\figas}[3]{Fig.\ref{fig:#1}(#2-#3)}
\newcommand{\figs}[2]{Figs.\ref{fig:#1}-\ref{fig:#2}}
\newcommand{\tab}[1]{Tab.\ref{tab:#1}}
\newcommand{\tabs}[2]{Tabs.\ref{tab:#1}-\ref{tab:#2}}
\newcommand{\tabb}[2]{Tabs.\ref{tab:#1} and \ref{tab:#2}}
\newcommand{\bom}{$\mu_{B}$~}
\newcommand{\eq}[1]{Eq.(\ref{#1})}
\newcommand{\eqs}[2]{Eq.(\ref{#1}-\ref{#2})}
\newcommand{\sct}[1]{Sec.(\ref{sec:#1})}

\newcommand{\sigmas}{$\sigma_{s}$~}
\newcommand{\sigmasa}{$\sigma_{s}^{*}$~}
\newcommand{\sigmab}{$\sigma_{pz}$~}
\newcommand{\sigmaab}{$\sigma_{pz}^{*}$~}
\newcommand{\pib}{$\pi_{p}$~}
\newcommand{\piab}{$\pi_{p}^{*}$~}
\newcommand{\nn}{N$_2$}
\newcommand{\bb}{B$_2$}

\newcommand{\ang}{\AA$^{-2}$~}
\newcommand{\mub}{$\mu_B$~}
\newcommand{\abit}{\emph{ab initio}~}
\newcommand{\abitt}{\emph{ab initio}~}
\newcommand{\insi}{\emph{in situ}~}
\newcommand{\abz}{A$_{BZ}$}
\newcommand{\etal}{\emph{et al.}}

\newcommand{\cel}{C$^\circ$}
\newcommand{\dg}{$^\circ$}

\newcommand{\rea}{X$^*$+Y$^*$$\rightarrow$XY$^*$}

\newcommand{\sta}{$^*$}
\newcommand{\airon}{$\alpha-$iron}

\newcommand{\nnrea}{2N$^*\rightarrow$N$_2^*$}
\newcommand{\bbrea}{B$^*$+B$^*\rightarrow$B$_2^*$}
\newcommand{\bnrea}{B$^*$+N$^*\rightarrow$BN$^*$}

\newcommand{\refcite}[2]{#1.[\onlinecite{#2}]} 
\newcommand{\refcites}[3]{#1.[\onlinecite{#2}-\onlinecite{#3}]} 
\newcommand{\simplecite}[1]{[\onlinecite{#1}]}

\newcommand{\blonski}{B\l\'{o}nski}



\setpagewiselinenumbers
\def\linenumberfont{\normalfont\small\sffamily}
\setlength\linenumbersep{0.1cm}
\title{Sub-monolayers of carbon on \airon~ facets: an \abit study}

\author{S. Riikonen,$^{1,2}$
A.~V. Krasheninnikov,$^{1,3}$
and R.~M. Nieminen$^{1}$}
\email{sampsa.riikonen@iki.fi}

\affiliation{$^1$ COMP/Department of Applied Physics, Helsinki University of
Technology, P.O. Box 1100, FI-02015, Finland}

\affiliation{$^2$Laboratory of Physical Chemistry, Department of Chemistry, Helsinki University of Technology
P.O. Box 55, FI-00014, Finland}

\affiliation{$^3$ Department of Physics, University of Helsinki,
P.O. Box 43, FI-00014,  Finland}

\begin{abstract}
Motivated by recent \insi studies of carbon nanotube growth from large transition-metal 
nanoparticles, we study various \airon~(ferrite) facets at different carbon 
concentrations using \abit methods.  The studied (110), (100) and (111) facets show
qualitatively different behaviour when carbon concentration changes.
In particular, adsorbed carbon atoms repel each other on the (110) facet,
resulting in carbon dimer and graphitic material formation.
Carbon on the (100) facet forms stable structures at concentrations of about 0.5 monolayer and at
1.0 monolayer this facet becomes unstable due to a frustration of the top layer iron atoms.
The stability of the (111) facet is weakly affected by the amount of adsorbed carbon
and its stability increases further with respect to the (100) facet with increasing carbon concentration.
The exchange of carbon atoms between the surface and sub-surface regions on the (111) facet is easier 
than on the other facets and the formation of carbon dimers is exothermic.  These findings are in accordance with a recent \insi experimental study where the existence of graphene decorated (111) facets is related to increased carbon concentration.
\end{abstract}



\maketitle 


\section{Introduction}
\label{sec:intro}
Carbon nanotubes (CNTs) are a versatile material with a wide range of potential technological applications
in fields such as mechanical engineering, electronics and biotechnology.
The chemical vapor deposition (CVD) method has established itself as the most effective
way to produce CNTs in mass quantities.  In this method, carbon containing molecules
(hydrocarbons, CO) are dissociated on catalytic nanoparticles (consisting typically of transition
metals and their alloys) where carbon eventually
forms graphitic structures and nanotubes.
For wider technological exploitation of CNTs, better control over the growth process is needed.
Controlling nanotube chirality and preventing particle poisoning and CNT growth termination
are the most important aspects.
In order to gain such a control, insight at microscopic level into the CNT growth process is 
needed.  This kind of insight can be obtained by performing \insi experiments where the CNT growth is directly observed.

Recently, several \insi environmental transmission-electron microscopy (TEM) studies of carbon fiber and nanotube growth have been carried out
 \cite{sharma04,helveg04,hofmann07,jensen05,abildpedersen06,begtrup09,rodriguezmanzo07,yoshida08,sharma09,rodriguezmanzo09}.
In these studies, the catalyst particles were either crystalline
and/or ``liquid-like'' (i.e. crystalline with high self-diffusivity) during the growth process.
Studied nanoparticle materials ranged from nickel\cite{helveg04,abildpedersen06,hofmann07,rodriguezmanzo09},
cobalt\cite{jensen05,rodriguezmanzo07,rodriguezmanzo09} and iron\cite{rodriguezmanzo07,begtrup09} to alloys\cite{rodriguezmanzo07} of these metals.
A common factor in many of these investigations is the appearance of step edges and new facets
as carbon is introduced to the nanoparticle and the growth of graphene layers from these special regions \cite{helveg04,abildpedersen06,rodriguezmanzo07,begtrup09}.
In the case of nickel, this phenomenon was attributed
to the stabilization of nanoparticle step edges upon carbon adsorption and the transport of catalyst metal atoms 
away from the step edge region.\cite{bengaard02,helveg04,abildpedersen06}
The energy barrier for carbon bulk diffusion in nickel was concluded to be very high when compared to any surface related diffusion phenomena. \cite{abildpedersen06}

In several studies, the dominant role of surface and sub-surface has been emphasized\cite{jensen05,abildpedersen06,hofmann07}
while other studies, considering mainly iron, suggest the importante of bulk 
diffusion\cite{rodriguezmanzo07,harutyunyan08,rodriguezmanzo09}.
Very recently, CNT growth from carbidic phase (cementite) in iron nanoparticles 
has been demonstrated\cite{yoshida08,sharma09}.

In a very recent study, solid-state \airon~nanoparticle was encapsulated inside multi-walled
carbon nanotubes (MWCNTs) while carbon was injected into the nanoparticle
by electromigration\cite{begtrup09}.  The nanoparticle was observed to stay solid and crystalline
during the growth.  Similar to earlier \insi studies,
new facets, showing growth of graphitic material, appeared on the nanoparticle surface.  
The orientation of the nanoparticle was analyzed and 
the MWCNT walls encapsulating the nanoparticle were parallel to the (110) facet.
The existence of the (111) facet was observed to depend strongly on carbon concentration, and
nanotube cap was formed on a rounded (100) facet.
\emph{Ab-initio} simulations were performed for ``graphenated'' and carbon saturated
surfaces in order to reproduce the nanoparticle shape\cite{begtrup09}.

In the case of \airon,~ the morphology of (110), (100) and (111) facets is quite different. This can result in very different diffusion barriers, carbon-carbon bond formation energetics and kinetics.
For better understanding of experiments it is important to 
perform \abit simulations and correlate computational results to the phenomena observed in the \insi studies. In particular, surface energies can be used to produce the physical shape of the nanoparticle which can be compared with the experiments\cite{barnard04,begtrup09}.  Activation energies of diffusion on
and into the facets may provide information about rate-limiting steps of graphitic material formation\cite{hofmann05,abildpedersen06}.

Recently, there have been several \abit studies related to these issues:
pure \airon~facets have been studied extensively by \blonski~and Kiejna\cite{blonski03,blonski07}, 
while carbon adsorption and diffusion on and into the (110) and (100) facets were studied by Jiang and Carter\cite{jiang05}.  The (100) facet has drawn some attention very recently, as segregated carbon atoms form stable,
periodic structures on this facet \cite{blum03,panaccione06,tan10}.  Some carbide surfaces have 
been studied with \abit methods by Chiou and Carter \cite{chiou03}.

In this work we study the \airon~(110), (100) and (111) facets at different carbon concentrations.
We address such topics as the interaction of adsorbate atoms in coadsorption configurations,
carbon diffusion (for the (111) facet) and formation of stable carbon-rich structures
(mainly on the (100) facet) in the topmost iron layer.
Relative surface energies as function of carbon concentration and energetics
of the smallest units involved in graphitic growth, the C$_2$ molecules, are studied.
This work is organized as follows: 
in \sct{methods}, simulation of iron-carbon systems, different carbon chemical potentials,
calculation of surface energies and computational details are discussed.
In \sct{results} the morphology of the studied \airon~facets is discussed followed
by the computational results.  Discussions and conclusions are made in \sct{conc}.

\begin{figure*}\centering
\putgraph{width=15cm}{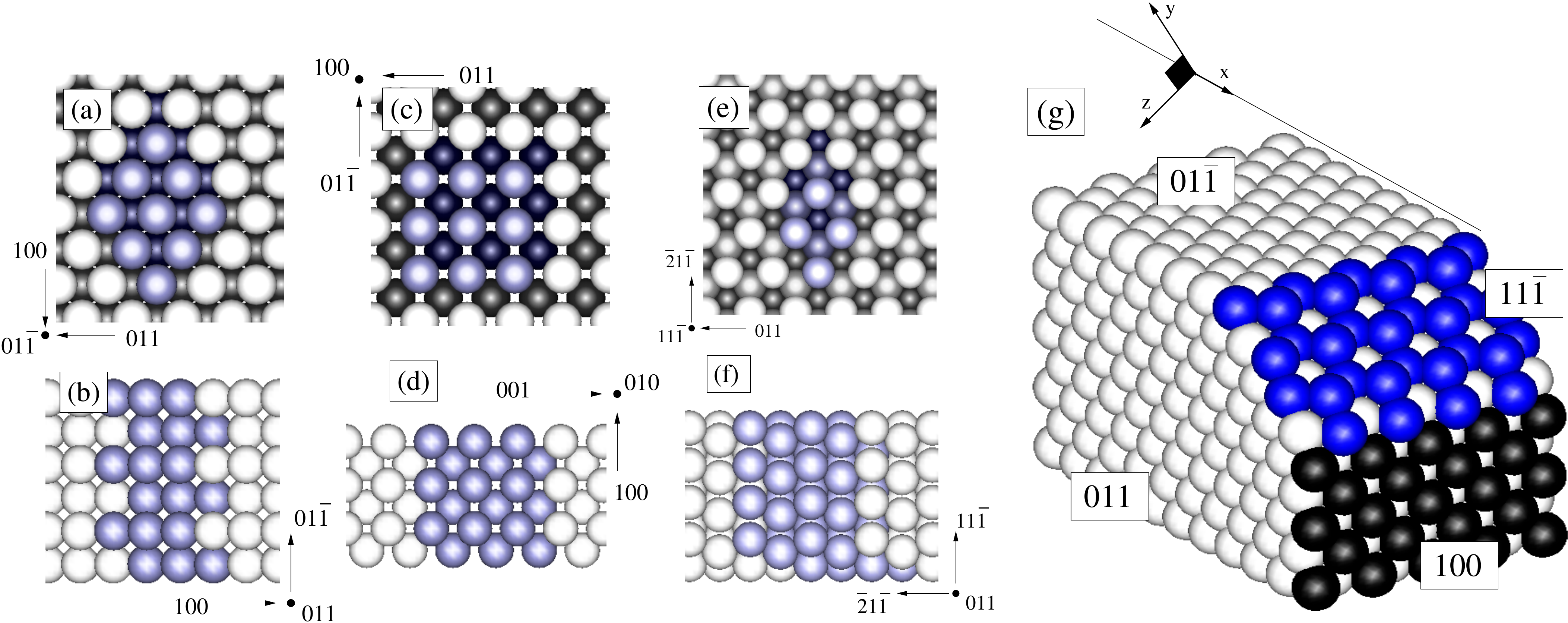}
\caption{(color on-line)
Computational unit cells for different \airon~surfaces used in the simulations
viewed from (a,c,e) top and from (b,d,f) side.
(a-b): (110), (c-d): (100) and (e-f): (111) surfaces.
Atoms in topmost (lowermost) layers are marked with brighter (darker) shades.
Atoms in the unit cells used in this work are marked with blue color.
(g) A portion of bulk $\alpha$-iron cut at different angles, demonstrating
the positions of different crystallographic surfaces.  For (100) surface, the topmost layer
is marked with black colour.  For (111) surface, the two topmost layers are marked with blue colour.
For (110) and (100) the unit cells correspond to 3$\times$3 periodicity, while for
(111) to 2$\times$2 periodicity.
\label{fig:unitcell}}
\end{figure*}

\begin{figure*}\centering
\putgraph{width=16cm}{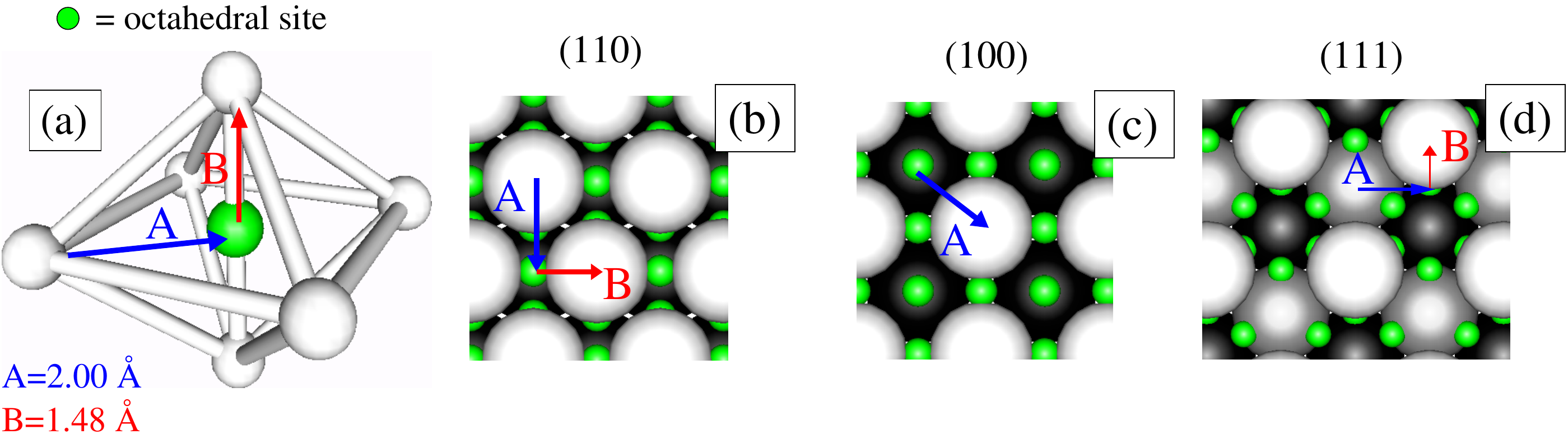}
\caption{(color on-line)
Local coordination at the bulk octahedral site in \airon~and how it is exposed on the
(110), (100) and (111) facets.  The octahedral site is marked in all insets
with green color. (a) The coordination of octahedral site in the bulk.
Coordination of octahedral sites on the (b) (110), (c) (100) and (d) (111) 
facets.
\label{fig:bcciron}}
\end{figure*}

\section{Methods}
\label{sec:methods}

\subsection{Iron-carbon systems}
\label{sec:ironsimu}
Iron with dissolved carbon exhibits a complex phase diagram as a function of temperature and carbon concentration, where \airon~(``ferrite'', bcc crystalline, ferromagnetic), $\gamma$-iron (``austenite'', fcc crystalline, antiferromagnetic) and cementite (Fe$_3$C)
are among the competing phases\cite{tilley}.  When considering nanoparticles instead of bulk, the phase
diagram will be modified; in particular, it is known that small iron nanoparticles prefer the $\gamma$-iron phase instead of the bulk \airon~ phase\cite{postnikov03,rollman07}.

While collinear spin calculations within the Density Functional Theory (DFT)
and General Gradient Approximation (GGA)\cite{gga} seem to work very well for \airon~\cite{jiang03}, 
it is not obvious how to perform calculations for $\gamma$-iron, as the
FCC iron, observed only at high temperatures, is paramagnetic.
These and some other problems have been discussed in detail by Jiang and Carter in \refcite{Ref}{jiang03}.

Considering systems involving nanoparticles, it has been shown that ``large'' iron nanoparticles 
(more than $\approx$ 100 atoms) prefer the bcc structure\cite{rollman07}.  This structure was
also observed in the large iron nanoparticle investigated in the recent \insi~experimental study\cite{begtrup09}.

\subsection{Technical Details}
In order to visualize the different \airon~ facets, a small volume of bulk, cleaved 
into several directions is illustrated in \figa{unitcell}{g}.  The form of this bulk
volume mimicks the elongated nanoparticles seen in the \insi experiment \cite{begtrup09}.
A real nanoparticle has, of course, several other facets which are not visualized in 
\figa{unitcell}{g} and not considered in this work.  Top and side views of the
facets are illustrated in panels (a-f) of \fig{unitcell}.

We employ the periodic supercell method to model the iron facets.
An infinite surface is modelled by a slab consisting of few layers of iron atoms,
with sufficient vacuum of 15\AA~between the slabs.  The slabs we have used are depicted
in \fig{unitcell} and they consist of 3$\times$3$\times$6 (the last number
denoting the number of layers) and 2$\times$2$\times$12 slabs for (110), (100) and (111)
surfaces, respectively.  Areas of the unit cells depicted in \fig{unitcell} are
51\AA$^2$, 72\AA$^2$, and 56 \AA$^2$ for (110), (100) and (111) slabs, respectively.

Next, in order to calculate adsorption and reaction energies on the surface, we
define a convenient energy quantity E$_s$ (shifted energies) as follows:
\begin{equation}\label{surfe}
E_s(X^*) = E(X^*) - E_0,
\end{equation}
where $E(X^*)$ is the energy of the adsorbed surface species X$^*$ and E$_0$ is the energy of
a surface unit cell without adsorbates.  Now the adsorption energy can be
written as follows:
\begin{equation}\label{ads}
E_{ads}=E(X^*)-E(X)-E_0 = E_s(X^*)-E(X),
\end{equation}
where $E(X)$ is the energy of an isolated atom in vacuum.

We also present adsorption energies, especially in the case of high carbon coverage, by using
some other chemical potential:
\begin{equation}\label{exc}
E_{c}=E(X^*)-(E_0+n\mu_c) = E_s(X^*)-n\mu_c,
\end{equation}
where $n$ is the number of carbon atoms and $\mu_c$ is the chemical potential. We
take the chemical potential as the energy per atom in graphene.  Energy E$_c$ then reflects the energy cost
to accommodate carbon atoms into the metal adsorbant instead of graphene.
In previous works\cite{jiang03,begtrup09,tan10} various chemical potentials 
have been used, including the energy per isolated carbon atom and the energy per carbon atom in graphite or in
graphene.

The shifted energy values of \eq{surfe} can be used to calculate reaction energetics on the adsorbate, i.e.
to look at energetics of processes like C$^*$+C$^*\rightarrow$C$_2^*$.  In the case of adsorbate-adsorbate
repulsion on the lattice, this can provide information about stress release upon dimer formation on the iron surface.
The energy for a reaction  X$^*$+Y$^*$$\rightarrow$XY$^*$ can be calculated as follows:
\begin{equation}\label{surfreact}
\Delta E = \big{(}E(XY^*)+E_0\big{)}-\big{(}E(X^*)+E(Y^*)\big{)}, \\
\end{equation}
This equation can be written, using the energy values E$_s$ of \eq{surfe} as follows:
\begin{equation}\label{surfreact2}
\Delta E = E_s(XY^*)-(E_s(X^*)+E_s(Y^*)).
\end{equation}
In the Results section, we tabulate values of E$_s$ and then use these tabulated values 
to calculate reaction energetics \rea~using \eq{surfreact2}.

The surface energy G of a specific nanoparticle facet with adsorbates
has in some earlier works\cite{barnard04_2,begtrup09} been defined as follows:
\begin{equation}\label{chapuza}
G = E(X^*)-NE^{bulk}-n\mu_c,
\end{equation}
where $N$ is the number of the surface atoms in the
slab used for simulations and $E^{bulk}$ is the energy per atom in the bulk.
As \eq{chapuza} reflects the energy cost to create a surface from the bulk
and adsorbing atoms on the surface, it is
more accurate to use an equation involving explicitly the surface energy $E_{surf}$ as follows:
\begin{equation}\label{surfes}
G = E_{surf}+E_{c},
\end{equation}
where E$_{c}$ is the adsorption energy of \eq{exc}.
In \eq{surfes} the surface energy $E_{surf}$ is a well defined quantity, while the chemical potential of the carbon adsorbates is sensitive to the source of carbon atoms.

In this work we use the values calculated
by \blonski~and Kjiena \cite{blonski03,blonski07} for the (110), (100) and (111) facets which have
been evaluated as discussed by Boettger\cite{boettger96}.
These are E$_{surf}$=140 meV/\AA~for the Fe(110) and Fe(100) surfaces and E$_{surf}$=160 meV/\AA~for 
the Fe(111) surface\cite{blonski07}.

Our calculations were performed in the framework of
the density functional theory (DFT), as implemented in the VASP code\cite{vasp2,vasp3}.
All calculations were done using projector-augmented waves (PAWs)\cite{paw} and the Perdew-Burke-Ernzerhof (PBE) 
generalized gradient approximation (GGA)\cite{gga}. We used the Monkhorst-Pack (MP) sampling\cite{mp} of the Brillouin zone in calculations involving the slab.  The sampling used was 7$\times$7$\times$1 in the case of all the slabs which corresponds to \abz$\approx$0.01 \AA$^{-2}$ (area in the reciprocal space per sampled k-point). A systematic search to find the optimal adsorption sites for C atoms and C$_2$ molecules was performed on the slabs of \fig{unitcell} along the lines of \refcite{Ref}{riikonen09}.

Spin polarization was included in all calculations. The cutoff energy of the plane wave basis set
was always 420 eV.  The mixing scheme in the electronic relaxation was the Methfessel-Paxton method\cite{mpax} of order 1.  Conjugate-gradient (CG) relaxation of the geometry was performed and if needed, the relaxation was continued with a
semi-Newton scheme.  This way we were able to reach a maximum force residual of $\approx$ 0.01 eV/\AA.
In all calculations the special Davidson block iteration scheme was used
and symmetries of the adsorption geometries were not utilized.  

As carbon chemical potential, we used either the energy of an isolated atom in vacuum
or the energy per atom in graphene.  For calculation of the chemical potential, identical parameters to those described
earlier in this section were used.  For an isolated, spin-polarized carbon atom calculated in a cubic unit cell with
15 \AA~sides, we obtained the total energy of E=-1.28 eV.  For graphene, and using a k-point sampling of 25$\times$25$\times$1
we obtained a lattice constant a=2.468\AA.  This is slightly larger than values obtained by LDA,\cite{wirtz04}
but is identical to a previous calculation using GGA\cite{gui08}.  The energy per carbon atom in graphene we 
obtained is E=-9.23 eV.

Nudged elastic band (NEB) calculations\cite{henkelman00} were performed with VASP. Atoms in the topmost layer were allowed to move freely, and in some cases, the atoms below the topmost layer were allowed to move into z-direction (normal to surface) only. Thus we were able to avoid the (artificial) collective movement of the surface slab atoms that sometimes occured during the minimization.

\section{Results}
\label{sec:results}

\subsection{Morphology of \airon~facets}
\label{sec:char}
We can expect from some earlier studies concerning carbon solution into bulk iron and adsorption on iron surfaces, \cite{jiang03,jiang05}
that carbon prefers sites of maximum coordination: in bulk iron, it 
prefers the 6-fold octahedral site\cite{jiang03} and on the (110) and (100) facets, carbon moves into sites
that offer highest possible number of iron neighbours\cite{jiang05}.
Keeping this in mind, we will give in this section a qualitative picture of carbon adsorption on
different facets.  This analysis is based on the octahedral site of bulk \airon.

The local coordination of the octahedral site is illustrated in \figa{bcciron}{a}.
When carbon is adsorbed into this site, a tetragonal distortion 
in the bcc lattice takes place and distance B is expanded.
As the bulk is cleaved along a specific direction in order to create a surface, the octahedra
become cleaved in a specific way, exposing octahedral sites.  
The way the octahedral sites are exposed in different facets, has been
illustrated in \figas{bcciron}{b}{d}.  On the (110) surface, the exposed octahedral sites
have neighboring iron atoms at distances A and B.  In the case of the (100) surface, there are several exposed
octahedral sites where the neighboring iron atoms are simply at a distance A.

Assuming that carbon tries to maximize its coordination with iron on the surface 
(as discussed above), it will always prefer 
an exposed octahedral site, as this kind of site offers maximum coordination within the bcc lattice.
The displacements of iron atoms surrounding a surface exposed octahedral should be very
similar to the bulk tetragonal distortion (i.e. expansion of (B) and a slight contraction of (A)). 
This distortion must be energetically very different on the distinct surfaces. Depending on how the distortion
of A and B fits the facet morphology, quite different adsorbate-adsorbate repulsions can be formed; for example, expanding B on the (110) facet (\figa{bcciron}{e}) consists of pushing neighboring iron atoms apart. On the (100) facet
there are many sites with no need to rearrange the iron atoms as only small contraction of A is needed.

\subsection{Bulk \airon~ facets}
\label{sec:purefer}
For the bulk iron lattice constant we obtained a=2.83~\AA, which agrees well
with an earlier computational value\cite{jiang03} and the experimental value of 2.87~\AA\simplecite{kittel}.
For the bulk magnetic moment we obtained  M=2.18 \mub.
In earlier works, interlayer relaxations for various \airon~ facets have been
studied\cite{blonski03,blonski07}.  The most important effect is the inward relaxation of
the outermost layer.
The interlayer relaxations can be sensitive to the slab size and to the scheme used (symmetric/non-symmetric
slab, number of fixed layers)\cite{blonski03,blonski07}.  In \tab{interlay} we compare our results
with previous ones.  In our scheme, the slab is non-symmetric as the atoms in the three bottom layers
are fixed.
In \refcite{Ref}{blonski03} three topmost layers were allowed to relax,
while in \refcite{Ref}{blonski07} freestanding slabs were considered.
As evident from \tab{interlay} we can see that the type of relaxation (either expansion or contraction) is
quite consistent.  Magnitudes of expansion/contraction have
a few differences of the order of 10\% in the case of the (111) facet, but on the other hand, 
in this facet the bulk interlayer distances are very small ($\approx$0.82\AA).

\begin{table}\small
\begin{tabular}{|lllllll|}
\hline
$d_{12}$ & $d_{23}$ & $d_{34}$ & $d_{45}$ & $d_{56}$  & $d_{67}$ &  \\
\hline\hline
\multicolumn{7}{|l|}{(110)} \\
\hline
-0.1 	 & 0.3      & -0.5     & -0.2     & 0.04      & 0.2	& \refcite{Ref}{blonski07}	\\
-0.11 	 & 1.16     & 1.14     &	  &	      &	        & \refcite{Ref}{blonski03}	\\
-0.4	 & 0.5	    & -0.7     &	  &	      & 	& This work \\
\hline\hline
\multicolumn{7}{|l|}{(100)} \\
\hline
-3.6 	 & 2.3 	    & 0.4      & -0.4     & -0.01      & -0.5   & \refcite{Ref}{blonski07}	\\
-3.09    & 2.83     & 1.93     &          &            &        & \refcite{Ref}{blonski03}	\\
-1.2	 & 3.4      & 3.5      &	  &	       & 	& This work \\
\hline\hline
\multicolumn{7}{|l|}{(111)} \\
\hline
-17.7    & -8.4     & 11.0     & -1.0     & -0.5       & 0.1    & \refcite{Ref}{blonski07}	\\
-6.74    & - 16.89  & 12.4     &          &            &        & \refcite{Ref}{blonski03}	\\
-3.8	 & -18.0    & 11.0     & 0.0      & -0.7      & 1.6 	& This work \\
\hline
\end{tabular}
\caption{
Interlayer relaxations in the slabs of \fig{unitcell} as percentage of the bulk distances.
$d_{ij}$ is the distance between layers i and j. 
\label{tab:interlay}}
\end{table}

\begin{table}\small
\begin{tabular}{|l|llllll|} 
\hline
     	& 1    & 2    & 3    & 4    & 5    & 6   \\ 
\hline\hline
110	& 2.57 & 2.28 & 2.17 & 2.17 & 2.29 & 2.56 \\ 
\hline
100	& 2.96 & 2.37 & 2.47 & 2.50 & 2.38 & 2.96 \\ 
\hline\hline
\multirow{4}{*}{111}
     	& 1    & 2    & 3    & 4    & 5    & 6   \\
\cline{2-7}
	& 2.89 & 2.39 & 2.49 & 2.30 & 2.28 & 2.15 \\ 
\cline{2-7}
	& 7   & 8    & 9    & 10   & 11   & 12 \\
\cline{2-7}
        & 2.20 & 2.24 & 2.29 & 2.46 & 2.29 & 2.84 \\
\hline
\end{tabular}
\caption{
Magnetic moment of atoms in different layers (\mub) in the slabs of \fig{unitcell}.  Number 1 denotes the bottom layer.
\label{tab:laymag}}
\end{table}
Values of magnetic moment in different layers are presented in \tab{laymag}.  Consistent with earlier
results\cite{blonski07} we observe that (100) has the highest top layer magnetic moment and that in all slabs, the value
of magnetic moment approach to that of bulk as we move inside the slab.
Our values for the moments in the topmost layer, 2.56\mub (110), 2.96\mub (100) and 2.84\mub (111),
compare well with the values of \refcite{Ref}{blonski07}, namely 2.59\mub (110), 2.95\mub (100) and 2.81\mub (111).

\subsection{Atomic carbon, coadsorption, dimers}
\label{sec:adgeom}
In this section we study atomic carbon adsorption \big{(}1/9 ML coverage for (110) and (100)
and 1/4 ML coverage for (111)\big{)} as well as dimer- and co-adsorption.
We study in detail how the adsorbates either repel or attract neighboring iron atoms
in the topmost iron layers and will use the resulting displacements of iron atoms as our leading argument when
describing the energetics at higher carbon concentrations in the next section.

\begin{figure*}\centering
\putgraph{width=12cm}{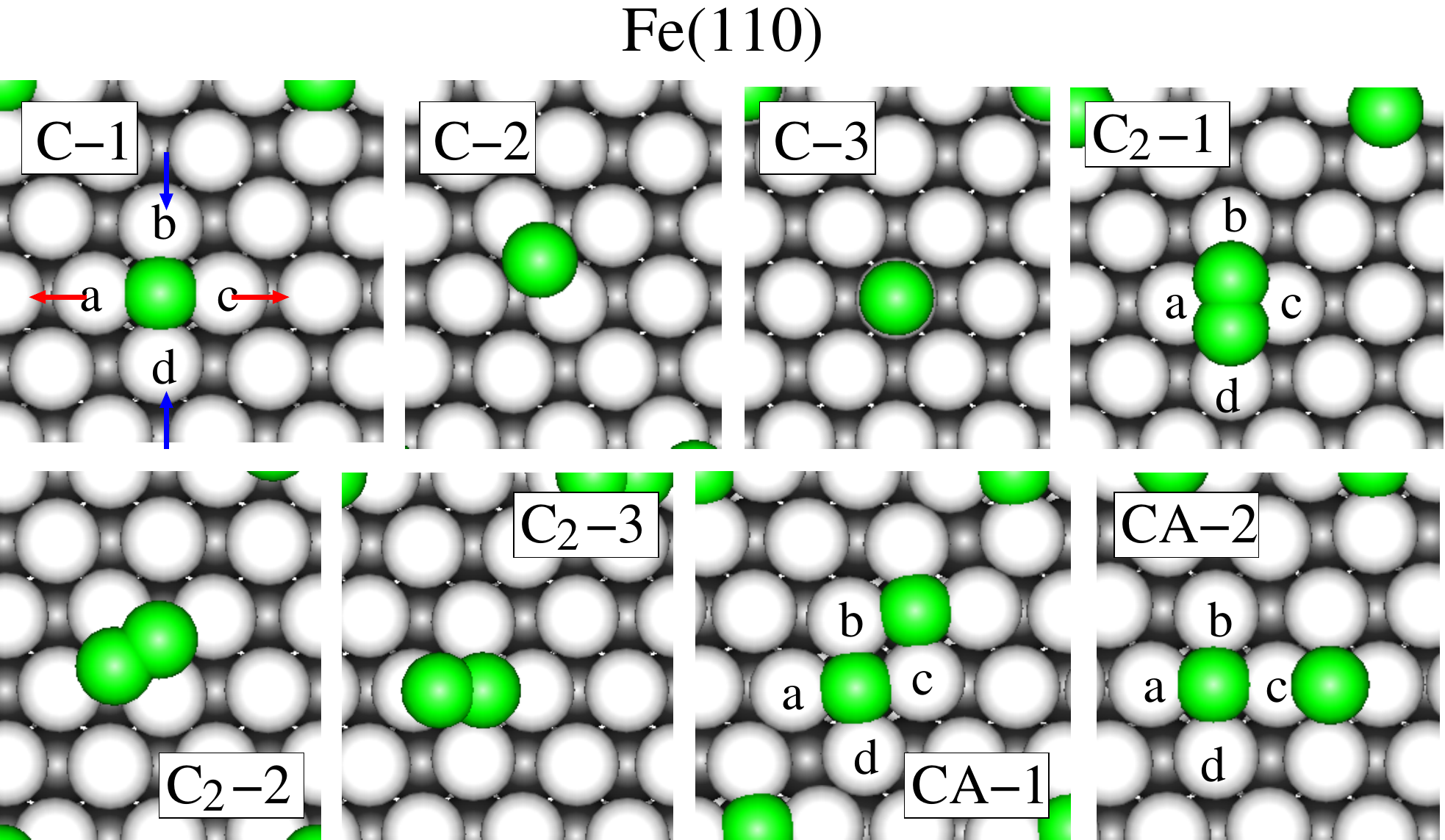}
\caption{
Some of the most stable geometries for C and C$_2$ on the Fe(110) surface.
Different geometries are tagged with the same labels as in \tab{surfnrjs}.
Coadsorption geometries, where atoms are adsorbed into the same
unit cell are tagged with the label ``CA''.  Iron atom displacements upon carbon adsorption in (C-1)
have been marked with arrows.  These can be related to \fig{bcciron}.
In (CA-2) one adsorbed carbon atom is pushed towards vacuum.
\label{fig:geoms1}}
\end{figure*}

\begin{figure*}\centering
\putgraph{width=12cm}{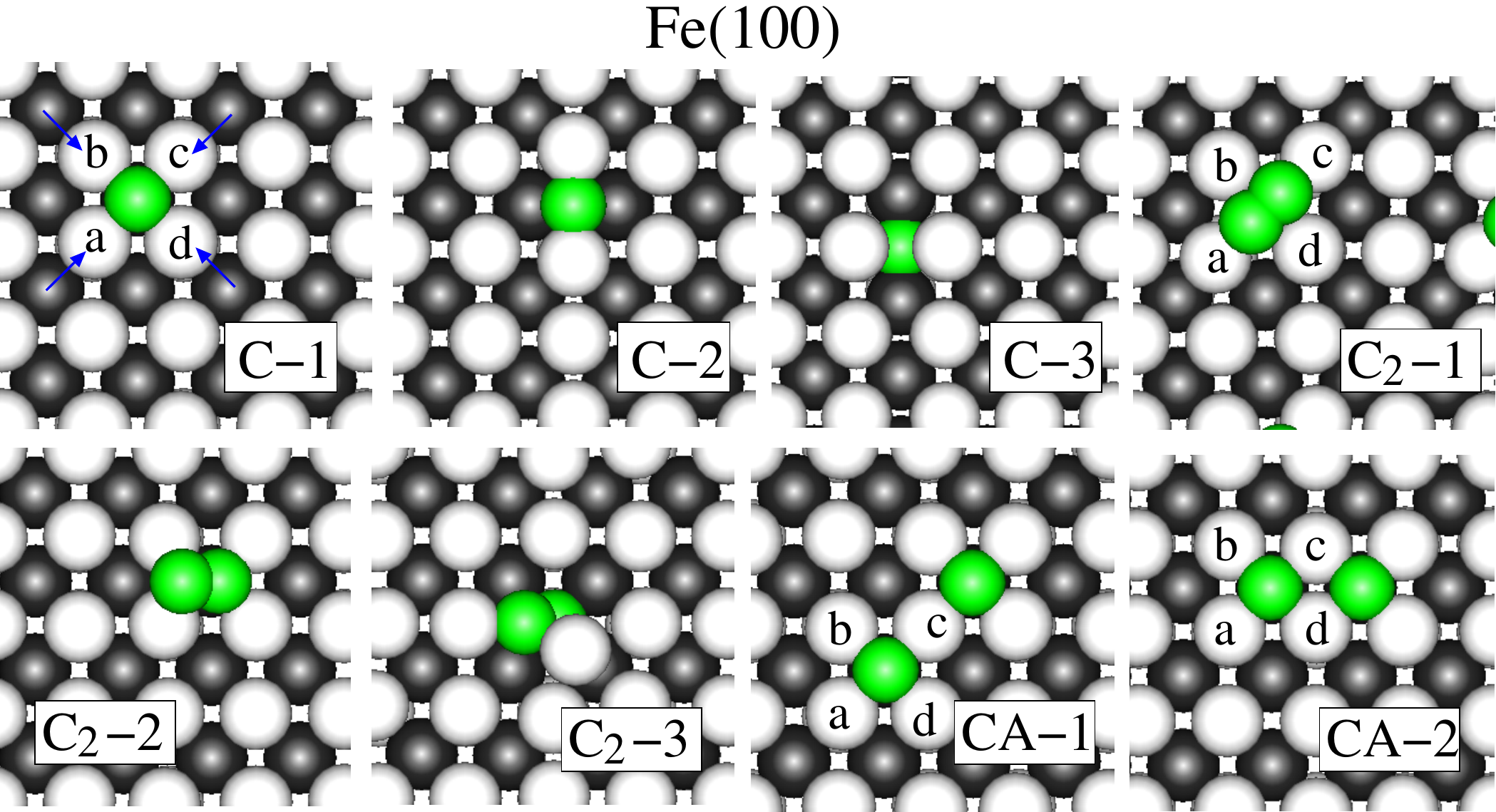}
\caption{
Some of the most stable geometries for C and C$_2$ on the Fe(100) surface.
Different geometries are tagged with the same labels as in \tab{surfnrjs}.
Coadsorption geometries, where atoms are adsorbed into the same
unit cell are tagged with the label ``CA''.  Iron atom displacements upon carbon adsorption in (C-1)
have been marked with arrows.  These can be related to \fig{bcciron}.
\label{fig:geoms2}}
\end{figure*}

\begin{figure*}\centering
\putgraph{width=13cm}{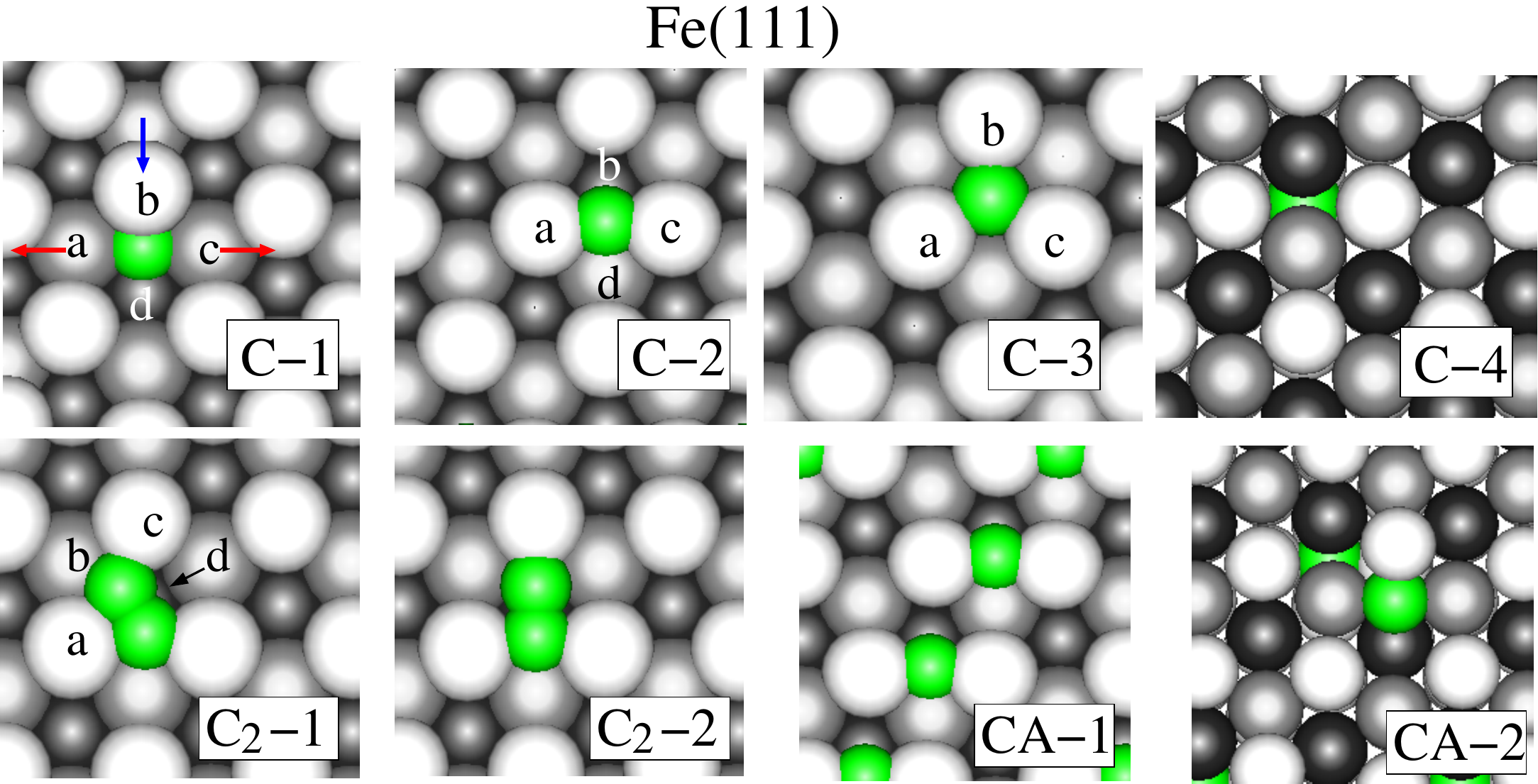}
\caption{
Some of the most stable geometries for C and C$_2$ on the Fe(111) surface.
Different geometries are tagged with the same labels as in \tab{surfnrjs}.
Coadsorption geometries, where atoms are adsorbed into the same
unit cell are tagged with the label ``CA''.  Iron atom  displacements upon carbon adsorption in (C-1)
have been marked with arrows.  These can be related to \fig{bcciron}.
\label{fig:geoms3}}
\end{figure*}

\subsubsection{Atomic carbon}
\label{sec:atomic}
Adsorption geometries for carbon atoms (C-1, C-2, etc.)
on the different facets are illustrated in \figs{geoms1}{geoms3}.
Corresponding energetics are tabulated in \tab{surfnrjs}.
\begin{table}
\begin{tabular}{|l|l|l|l|l|}
\hline
Adsorbate & E$_{ads}$ (eV) & E$_c$ (eV) & E$_s$ (eV) & BL (\AA) \\
\hline\hline
\multicolumn{5}{|l|}{Fe(110)} \\
\hline
C-1&-7.98&-0.03&-9.26&\\
C-2&-6.91&1.04&-8.19&\\
C-3&-5.48&2.47&-6.76&\\
C$_2$-1&-8.19&0.73&-17.72&1.35 (1.31)\\
C$_2$-2&-8.04&0.88&-17.57&1.38\\
C$_2$-3&-7.12&1.81&-16.64&1.32\\
CA-1&&0.2&-18.26&\\
CA-2&&0.74&-17.71&\\
\hline\hline
\multicolumn{5}{|l|}{Fe(100)} \\
\hline
C-1&-8.45&-0.5&-9.73&\\
C-2&-7.2&0.74&-8.48&\\
C-3&-7.18&0.77&-8.46&\\
C$_2$-1&-8.26&0.67&-17.79&1.33\\
C$_2$-2&-7.91&1.01&-17.44&1.36\\
C$_2$-3&-7.51&1.42&-17.03&1.44\\
CA-1&&-1.02&-19.48&\\
CA-2&&-0.92&-19.38&\\
\hline\hline
\multicolumn{5}{|l|}{Fe(111)} \\
\hline
C-1&-7.74&0.2&-9.02& \\
C-2&-7.69&0.26&-8.96& \\
C-3&-7.43&0.52&-8.71& \\
C-4&-7.41&0.53&-8.69&\\
C$_2$-1&-8.95&-0.02&-18.47&1.40 \\
C$_2$-2&-8.87&0.05&-18.4&1.38\\
CA-1& &0.61&-17.84& \\
CA-2& &0.82&-17.63& \\
\hline
\end{tabular}
\caption{
Adsorption energies E$_{ads}$ (see \eq{ads}), energies E$_c$ (see \eq{exc}) and shifted energies E$_s$ (see \eq{surfe}).  Values of E$_s$  can be used directly to calculate reaction energies on the surface by using \eq{surfreact2}.  Values for C atoms and C$_2$ molecules in different adsorption geometries on the $\alpha$-iron (110), (100) and (111) surfaces have been tabulated.  Bond lengths (BL) for adsorbates and in vacuum (in parenthesis) are listed.  Sites and geometries have the same labels as in \figs{geoms1}{geoms2} and in \tab{surfreact}.  Coadsorption geometries are tagged with the label ``CA''.
\label{tab:surfnrjs}
}
\end{table}
Taking a closer look at \figs{geoms1}{geoms3}, we can see that carbon favors
the exposed octahedral sites of \fig{bcciron} as discussed in \sct{char}.

The most favorable (C-1) sites for (110) and (100) are consistent with previous calculations\cite{jiang05,begtrup09,tan10}.
From \tab{surfnrjs} the adsorption energies are
-7.98 eV and -8.45 eV for the (110) and (100) surfaces, respectively.
These compare well with earlier computed
values of -7.92 eV\simplecite{jiang05} and -7.97 eV\simplecite{begtrup09} for (110) and with
-8.33 eV\simplecite{jiang05}, -8.335 eV\simplecite{tan10} for the (100) surface.
 
The iron atom below the adsorbed
carbon atom on 100/C-1 shifts downwards, corresponding to expansion
of (B) in \fig{bcciron}.  The coordination of carbon on (100) is fivefold\cite{jiang05,tan10} and
it is bonded to the iron atom below at a distance of $1.98$ \AA.
In the following we analyze in more detail the intralayer relaxations in the topmost surface layer,
using as a guide the qualitative discussion of these relaxations made in \sct{char}; this kind of
analysis, based on the tetragonal distortion of bulk adsorption, was made to some extent in 
\refcite{Ref}{jiang05}, but only for the case of sub-surface adsorption.

In 110/C-1 of \fig{geoms1} there are considerable intralayer relaxations in the topmost layer.
The distance between iron atoms (b) and (d) contracts by 5 \% 
(0.20\AA,~corresponding to contraction of A in \fig{bcciron})
while the distance between (a) and (c) expands 23\% (0.65\AA,~corresponding to expansion of B).
In 100/C-1 of \fig{geoms2} the displacement of iron atoms is smaller:  now both distances (bd) and (bc) contract only by 5 \% (0.15\AA), corresponding simply to the slight contraction of A;
due to the specific cleaving of the octahedron in \fig{bcciron}, the Fe(100) surface offers 
exposed octahedral sites for carbon with very little need to move the surrounding iron atoms.

From \tab{surfnrjs} we can see that adsorption into 100/C-1 is 0.47 eV more favorable
than adsorption into 110/C-1 (similar to the value of 0.41 eV obtained in \refcite{Ref}{jiang05}).
When looking at energies E$_c$, we observe that 
at a low coverage of 1/9 ML (i.e. single adsorbed atom), it is more favorable for the carbon atom to be adsorbed into the iron surface than to be incorporated into graphene.  While in the case of (110), this tendency is very weak (only $\sim$ 30 meV), for (100) it is more significant (0.5 eV).  As will be discussed below, E$_c$ is very sensitive to the amount of carbon adsorbed on the facets: at lower concentrations than we are considering in this paper (less than 1/9 ML), E$_c$ should clearly become negative also for (110).

The remaining adsorption geometries for (110), i.e. 110/C-2 and 110/C-3 are metastable down to $\approx$ 0.012 meV/\AA~and they lie more than 1 eV higher in energy than 110/C-1.  Their characteristics (local minimum, higher order
saddle point, etc.) have been discussed in more detail in \refcite{Ref}{jiang05}..

Comparing \figa{bcciron}{d} and the optimal adsorption sites of carbon atoms in \fig{geoms3}
we can see that carbon prefers exposed octahedral sites on the (111) facet.
In 111/C-1, there is a 0.74\AA~expansion in the distance of atoms (b-d) and a 
0.2\AA~contraction in the distance of atoms (a-c), corresponding again to (B) and (A) in
(\figa{bcciron}{a}).  While adsorbate 111/C-2 exhibits very similar distortions,
111/C-3 breaks the trend a bit as it does not adsorb into an exposed octahedral site.
It finds a high coordination by moving atoms (a),(b) and (c) instead.
Atoms (a),(b) and (c) all move symmetrically $\sim$0.3\AA~and their distance to the carbon
atoms becomes 2.1\AA.  There is also one iron atom directly below the carbon at a
distance of 1.85\AA.  Adsorbate 111/C-4 is simply a carbon atom adsorbed at a bulk-like
octahedral site.

When looking at the adsorption geometry C-1, we can observe that it is by definition a ``sub-surface''
site, i.e. the carbon atom resides below the topmost iron layer.  On the other hand, it
has not yet obtained a coordination with surrounding iron atoms similar to that in bulk.  On the contrary,
C-2 is clearly a ``surface'' adsorption site.  The energy difference between C-1 (a ``semi'' sub-surface site)
and C-2 (``surface'' site) is minimal, only 50 meV.

%
The energetics for carbon adsorption on (111) in \tab{surfnrjs} are not directly comparable
to those reported in \refcite{Ref}{begtrup09}, as in that work the motion of iron atoms was
constrained (some of the sites will have diffent local geometries upon relaxation).
Our values for the C-1 (-7.74 eV) and C-3 (-7.43 eV) sites
are very close to the value reported in \refcite{Ref}{begtrup09} (-7.60 eV) for a similar site.

\subsubsection{C$_2$ dimer}
Assuming that carbon atoms in the dimer prefer similar high-coordinated sites
as the individual atoms while maintaining a reasonable carbon-carbon bond length, there
are not good possibilities to achieve this on the (110) and (100) surfaces, as the
optimal C-1 sites lie far away from each other.  The case of (111) is very different;
looking at \figa{bcciron}{d} we can see that there is an abundance of optimal adsorption 
sites within the bond length of a carbon dimer.
We can then expect that C$_2$ dimer is most stable on the (111) surface.

The optimal C$_2$ adsorption geometries on the (110) and (100)
of \figs{geoms1}{geoms2} are in both facets quite similar:
individual carbon atoms are 1-3 fold coordinated to iron:
one Fe-C distance in both cases is 1.85\AA, while the remaining two Fe-C
distances are $\approx$ 2.0\AA.
The C-C bond length for 110/C$_2$-1 
is slightly expanded while for 100/C$_2$-1 it is closer
to the isolated C$_2$ bond length.
In 110/C$_2$-1, both iron atom distances ac and bd expand by $\sim$5\%
while in 100/C$_2$-1, the distance between a and c is expanded by $\sim$15\%.
Of the remaining C$_2$ adsorption geometries
110/C$_2$-2 exhibits similar trend as C$_2$-1: both C atoms reside in a site $\sim$3-fold to iron atoms.
Other C$_2$ adsorption geometries are trying to adopt positions where the adsorbed carbon
can reside in C-1, C-2 or C-3 sites

The optimal position, C$_2$-1 for the carbon dimer on (111) is depicted in \fig{geoms3}.
Both carbon atoms reside in a C-1 site while the Fe-C distances are $\approx$2 \AA.
The C-C length in the dimer is expanded by $\approx$6\%.
The adsorbate C$_2$-2 exhibits a similar trend.
Comparing the adsorption energies of C$_2$-1 in the case of the different facets
(\tab{surfnrjs}), the adsorption of the carbon dimer on 111
is at least $\approx$ 0.7 eV more favorable than on the other facets; as discussed above,
this is because the (111) facet offers nearby optimal adsorption sites for the carbon atoms.

The reaction energetic of \tab{surfreact} further demonstrate
that it is much more favorable to form carbon dimers on the (111) facet 
than on the (110) and (100) facets.   The (100) facet favors less dimer formation than 
the other ones, as having carbon in atomic form is energetically most favorable on this facet.

\subsubsection{Coadsorption}
\label{sec:coad}
We have studied coadsorption configurations by placing two carbon atoms
in a sublattice of the most optimal adsorption sites of the individual carbon atoms.
The optimal coadsorption sites we found are marked with tags ``CA'' in \figs{geoms1}{geoms3}.

In the case of (110) (\fig{geoms2}) the two coadsorption configurations considered become very different;
in 110/CA-1, both atoms reside in a 110/C-1 adsorption site.  However, in 110/CA-2, one of the atoms is forced
to move from the C-1 site towards vacuum.  As discussed earlier, this results from the expansion of 
distance (B) (\fig{bcciron}) on the (110) facet;
in 110/C-1, atoms (a) and (c) are pushed apart and in 110/CA-2 the carbon atoms are pushing
the same atom (c) into opposite directions.  This creates strong repulsion between the carbon atoms
and only one carbon atom can be accomodated into the C-1 site.
This results in a notable, 0.5 eV energy difference between the 110/CA-1 and 110/CA-2 configurations

The situation is very different on the (100) surface.  As discussed above, the adsorption of carbon 
to the 100/C-1 site does not involve considerable motion of the surface iron atoms, as the only displacement
needed is the very small contraction of (A) (\fig{bcciron}).  
In the optimal coadsorption geometries 100/CA-1 and 100/CA-2 of \fig{geoms2} there are indeed very minor displacements
of iron atoms towards the adsorbed carbon.
In CA-2 both carbon atoms attract iron atoms (c) and (d), while in CA-1 they pull only one common iron atom (c).
In \tab{surfnrjs} we can see that this results in a small 0.1 eV energy difference between CA-1 and CA-2.

In the case of the (111) surface, there are quite many neighboring optimal
adsorption sites (C-1 and C-2) and so the number of coadsorption configurations becomes large.  The two most
optimal configurations we found (CA-1 and CA-2) are illustrated in \fig{geoms2}.
As evident from \tab{surfnrjs}, their energy difference is only 0.2 eV.

\begin{table*}\small
\begin{tabular}{lll}

\begin{tabular}{|l|l|}
\hline
\multicolumn{2}{|l|}{Fe(110)} \\
\hline\hline
Reaction & $\Delta E$ \\
\hline
2 (C-1) $\rightarrow$ C$_2$-1&0.79\\
2 (C-2) $\rightarrow$ C$_2$-1&-1.35\\
(C-1) + (C-2) $\rightarrow$ C$_2$-1&-0.28\\
\hline
CA-1 $\rightarrow$ C$_2$-1&0.53\\
CA-2 $\rightarrow$ C$_2$-1&-0.01\\
\hline
2 (C-1) $\rightarrow$ CA-1&0.26\\
2 (C-1) $\rightarrow$ CA-2&0.81\\
\hline
\end{tabular}

&

\begin{tabular}{|l|l|}
\hline
\multicolumn{2}{|l|}{Fe(100)} \\
\hline\hline
Reaction & $\Delta E$ \\
\hline
2 (C-1) $\rightarrow$ C$_2$-1&1.67\\
2 (C-2) $\rightarrow$ C$_2$-1&-0.82\\
(C-1) + (C-2) $\rightarrow$ C$_2$-1&0.43\\
\hline
CA-1 $\rightarrow$ C$_2$-1&1.69\\
CA-2 $\rightarrow$ C$_2$-1&1.59\\
\hline
2 (C-1) $\rightarrow$ CA-1&-0.02\\
2 (C-1) $\rightarrow$ CA-2&0.08\\
\hline
\end{tabular}

&

\begin{tabular}{|l|l|}
\hline
\multicolumn{2}{|l|}{Fe(111)} \\
\hline\hline
Reaction & $\Delta E$ \\
\hline
2 (C-1) $\rightarrow$ C$_2$-1&-0.42\\
2 (C-2) $\rightarrow$ C$_2$-1&-0.54\\
(C-1) + (C-2) $\rightarrow$ C$_2$-1&-0.48\\
\hline
CA-1 $\rightarrow$ C$_2$-1&-0.63\\
CA-2 $\rightarrow$ C$_2$-1&-0.84\\
\hline
2 (C-1) $\rightarrow$ CA-1&0.21\\
2 (C-1) $\rightarrow$ CA-2&0.41\\
\hline
\end{tabular}

\\

\end{tabular}

\caption{
Reaction energies for forming C$_2$ dimers and for bringing carbon atoms into coadsorption
configurations.  Geometries are tagged with the same labels (C-1, C-2, etc.) as in
\tab{surfnrjs} and \figs{geoms1}{geoms2}.
Reaction energies are calculated by taking the corresponding energies 
E$_s$ from \tab{surfnrjs} and applying \eq{surfreact2}.
\label{tab:surfreact}
}
\end{table*}

\subsection{Higher carbon concentrations}
\label{sec:for}

\begin{figure*}
\putgraph{width=15cm}{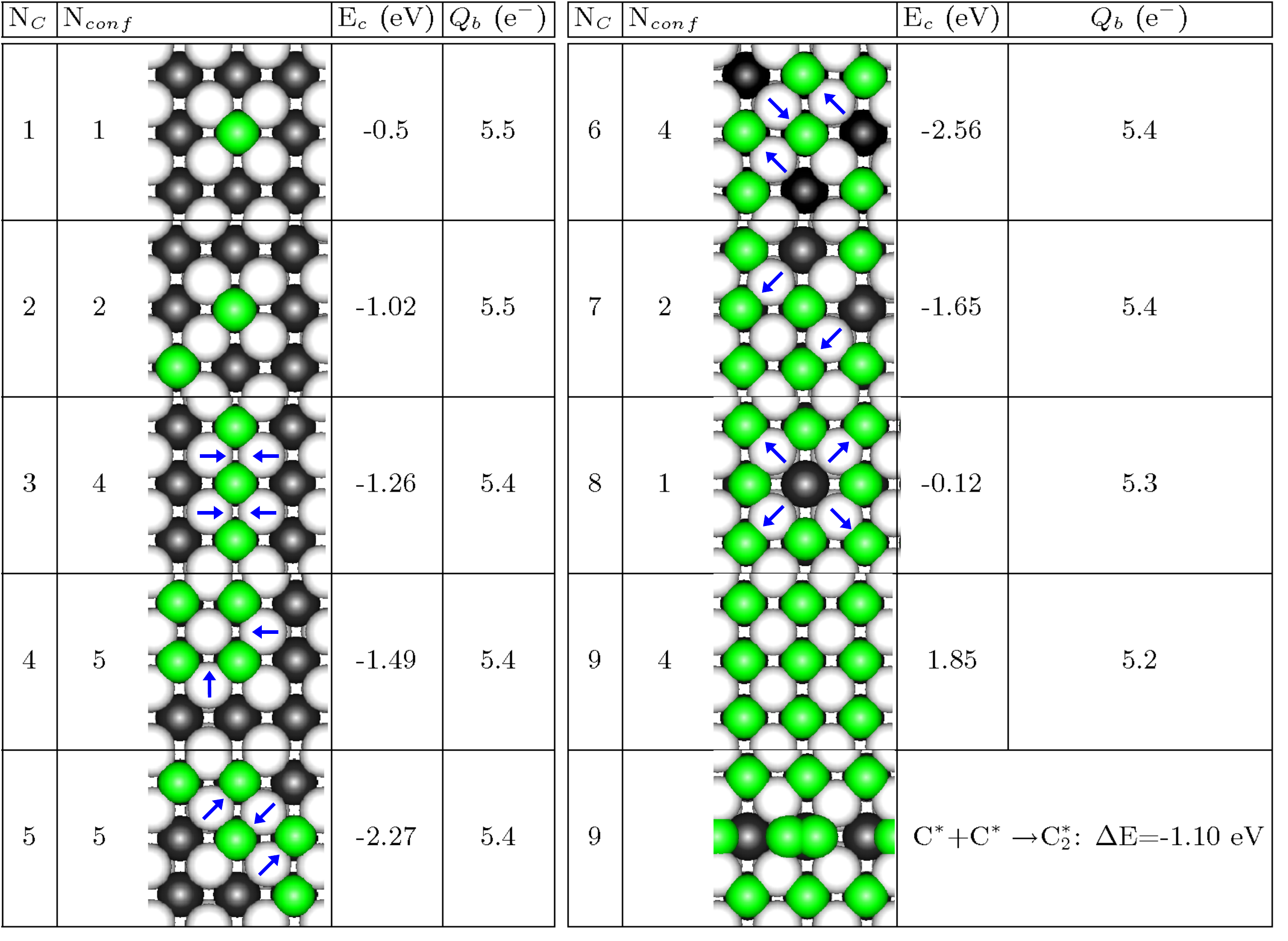}
\caption{
Adsorption geometries where carbon atoms are adsorbed in the sublattice
formed by the C-1 sites (see \fig{geoms2}).
N$_C$ is the number of carbon atoms and N$_{conf}$ the number of all possible coadsorption configurations
when N$_C$ atoms are adsorbed on a (3$\times$3) C-1 sublattice.
For a particular coadsorption geometry illustrated in the table, energy E$_c$ and
the mean Bader electron occupations for carbon atoms ($\bar{Q}_b$) have been tabulated.
Displacements of some iron atoms upon carbon adsorption have been highlighted by blue arrows.
In the last row and column, the energy gain when forming a C$_2$ molecule at
the 1 ML concentration is calculated.
\label{fig:highdens}
}
\end{figure*}

\begin{figure*}
\putgraph{width=12cm}{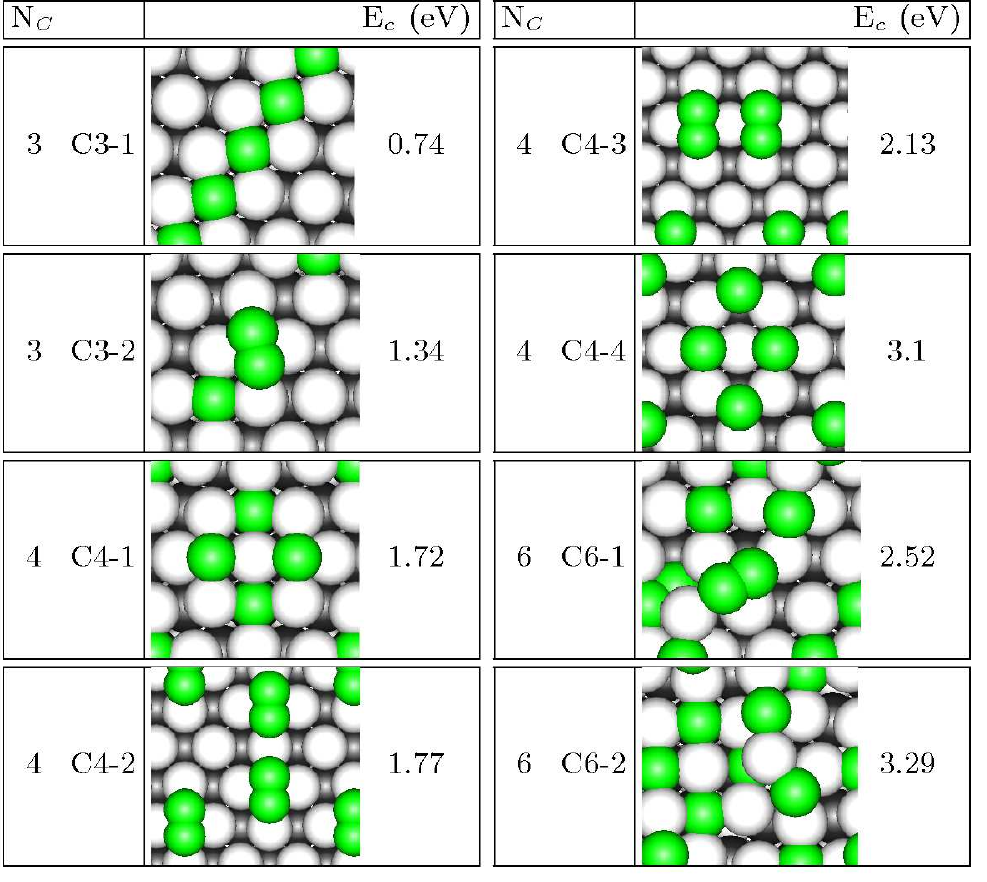}
\caption{
Relaxed adsorption geometries on (110) after carbon atoms have been placed in a sublattice
formed by the C-1 (C3-1, C4-1,C6-2), C-3 (C3-2, C4-4, C6-1) and C$_2$-1 (C4-2, C4-3) configurations of
\fig{geoms1}. Number of carbon atoms per unit cell (N$_C$)
and energies E$_c$ have been tabulated.
\label{fig:highdens2}
}
\end{figure*}

\begin{figure*}
\putgraph{width=11cm}{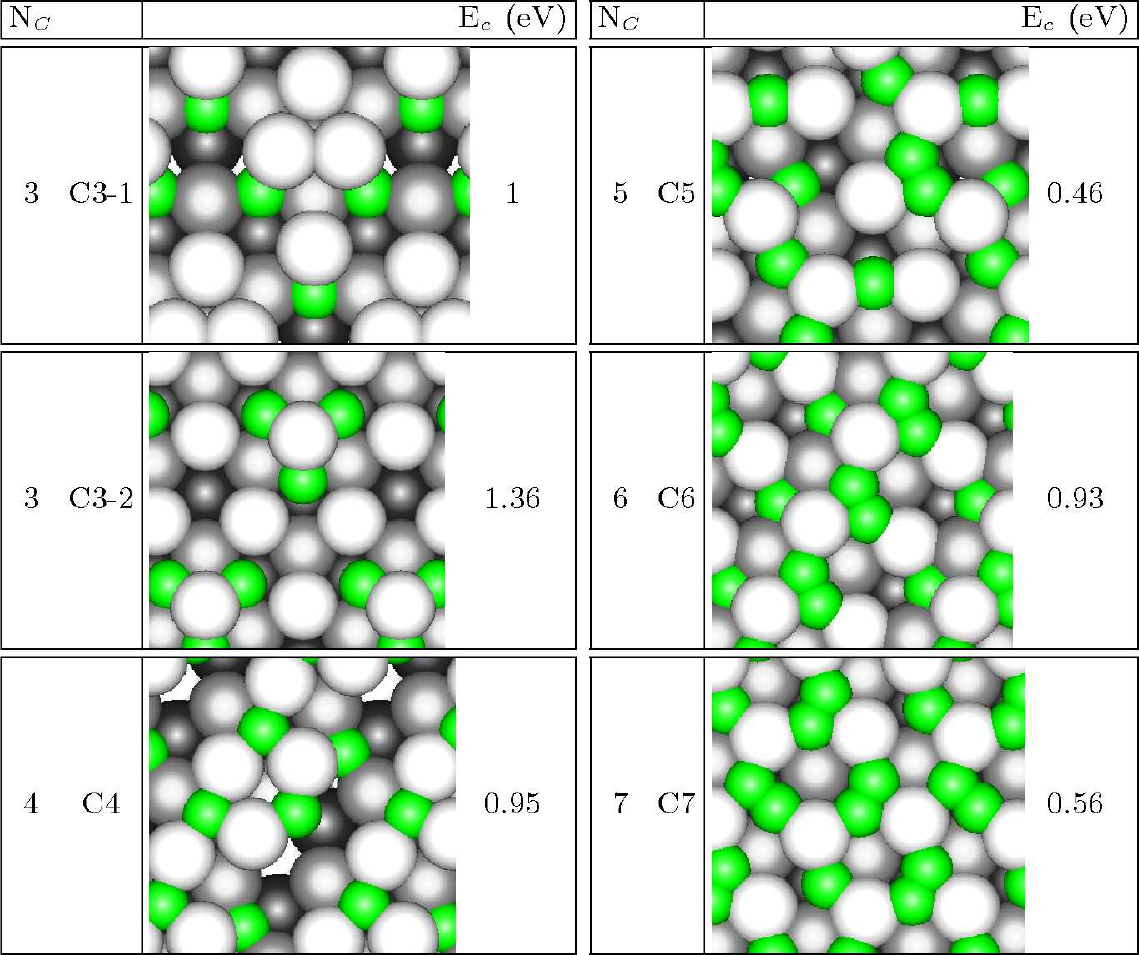}
\caption{
Relaxed adsorption geometries on (111) after carbon atoms have been placed in a sublattice
formed by the C-1 and C-2 sites (see \fig{geoms3}).  Number of carbon atoms per unit cell (N$_C$)
and energies E$_c$ have been tabulated.
\label{fig:highdens3}
}
\end{figure*}

\begin{figure*}\centering
\putgraph{width=12cm}{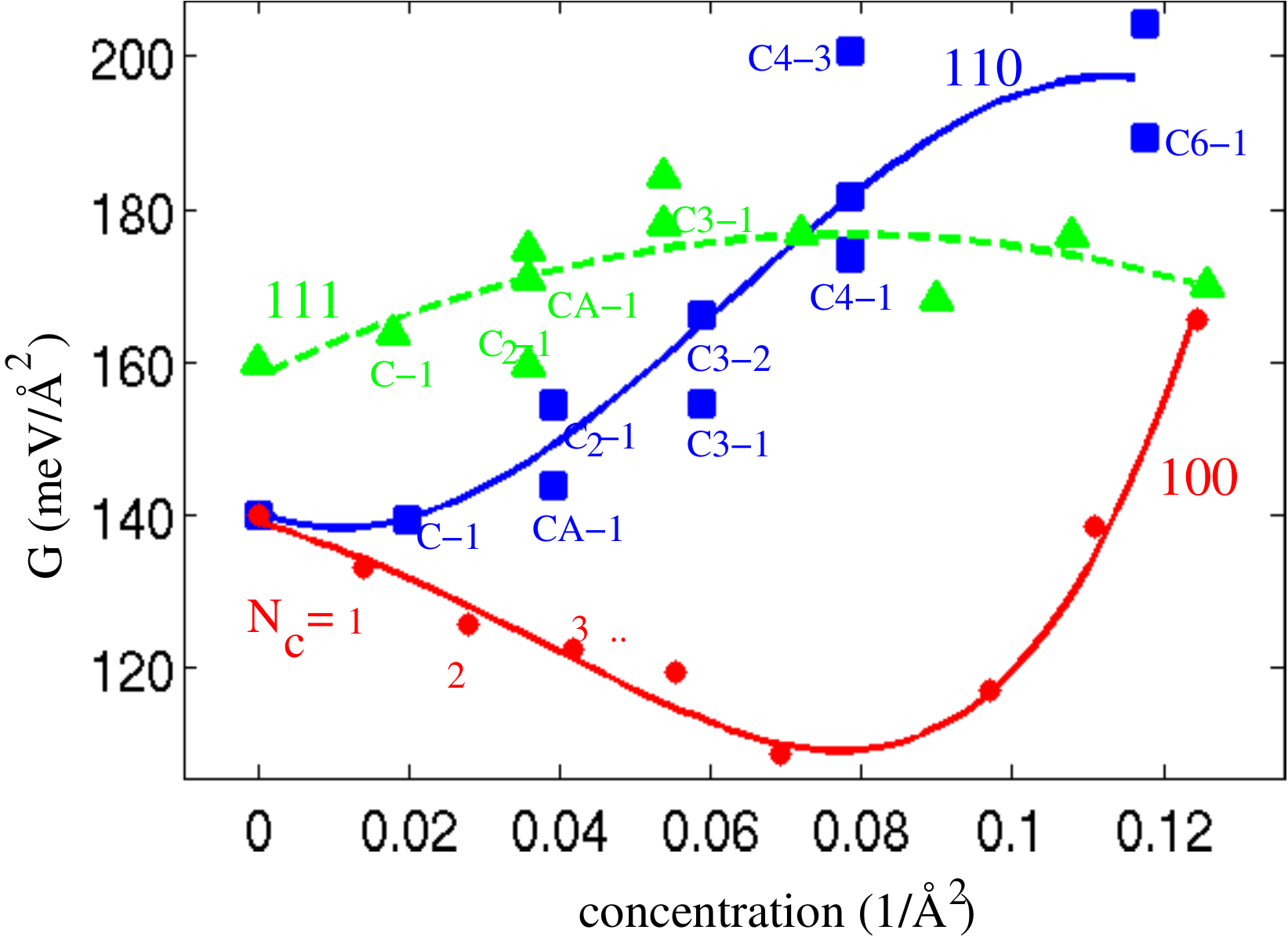}
\caption{
Surface energy G (\eq{surfes}) as function of carbon coverage (atoms per \AA$^2$) on the (110), (100) and (111) facets.
Energies correspond to the fully relaxed adsorption geometries of \figs{geoms1}{highdens3}.
Energy at zero concentration has been fixed to the surface energy of the facet \big{(}140 meV/\AA$^2$ for (110) and (100), 160 meV/\AA$^2$ for (111)\big{)}, as described by \eq{surfes}.  
\label{fig:carboexes}}
\end{figure*}

In this section, we study the effect of high carbon concentrations 
\big{(}$>$2/9 ML for (110) and (100), $>$3/4 ML for (111)\big{)} on energies E$_c$ (\eq{exc}).
Carbon is adsorbed on the sublattice formed by the most favorable C-1 adsorption
sites.
Our study is most systematic for the (100) surface as there is an obvious
way how to place an increasing number of carbon atoms on the surface: we do not expect 
considerable displacement of iron atoms from the bulk positions nor the displacement
of carbon atoms away from the optimal C-1 adsorption sites.
On the other hand, as we saw in previous sections, on (110) surface
adsorbed carbon atoms start to repel each other, while on (111) we can expect dimer formation.
Our objective for (110) and (111) is simply to further demonstrate these points 
(adsorbate repulsion, dimer formation) at higher carbon concentrations.
We start with the case we studied most systematically, i.e. with the (100) surface.

\textbf{(100) surface.}~
Carbon concentration of up to 1 ML has been considered for the (100) facet 
in \fig{highdens}.  A minimum of energy E$_c$ as function of carbon concentration
occurs at the coverage of 6/9 = 0.667 ML 
at an adsorption configuration ``3$\sqrt(2)\times\sqrt(2)$''
that has been reported earlier in an experimental\cite{panaccione06}
and theoretical\cite{tan10} study; in this adsorption configuration, carbon atoms
form an infinite, two atom wide ribbon on the C-1 sublattice.
There exists even more favorable coadsorption configuration
``c(2$\times$2)'' which takes place at the coverage of 0.50 ML\cite{blum03,jiang05,tan10}.
Both of these stable structures have been analyzed using STM and/or LEED\cite{blum03,panaccione06}
For the 6/9 ML coadsorption structure presented in \fig{highdens}
we observe same phenomena as reported in an earlier study\cite{tan10},
most notably the displacements of iron atoms in the topmost layer as illustrated schematically
in \fig{highdens}.

The most interesting phenomena in the case of the (100) surface is the behaviour 
of E$_c$ in \fig{highdens} as function of carbon concentration: the energy lowers
as more carbon is adsorbed, implying the existence of stable iron-carbon phases, while
at increased $>$0.5 ML carbon concentrations the situation becomes unstable and graphene formation is favored.
This behaviour can be explained by considering the displacements of iron atoms in the topmost surface layer.

From the reconstruction patterns presented by arrows in \fig{highdens}
we can see that iron atoms move always towards the adsorbed carbon. These changes in iron atom positions are typically
in the range of $\sim$0.2-0.3\AA~and, this is in accordance what was discussed in earlier sections.
As carbon concentration increases, iron atoms become increasingly ``frustrated'' as they are surrounded by 
carbon; at 1.0 ML, an iron atom has so many neighboring carbon atoms that it cannot obtain optimal bonding conditions
with any one of them.

This frustration mechanism results in sudden change in the energetics on the (100) facet;
even at such a high concentration as 8/9 ML, it is more favorable
to adsorb carbon atoms to the C-1 sublattice than incorporate them into graphene, but when going from
8/9 ML to 1 ML, this tendency is suddenly reversed.
In the last row and column of \fig{highdens}, we present the energy gain when releasing 
surface frustration by dimer formation.  It is quite large, over 1 eV.

\textbf{(110) surface.}~
Some configurations of carbon for up to 1 ML have been considered for the (110) facet in \fig{highdens2}.
The initial configurations consisted of adsorbing carbon atoms on either the C-1,
C-3 or C$2$-1 sublattice (see \fig{geoms1}) and the final, relaxed geometries are presented in \fig{highdens2}.
As we discussed in \sct{coad}, at 2/9 ML coverage there are coadsorption configurations where carbon
adsorbates start to repel each other.  Such effects become more important at
higher carbon concentrations.  In \fig{highdens2} and for a 
3/9 ML concentration, there is still one possibility to adsorb carbon without 
creating an excess of surface stress, similar to CA-1 of \fig{geoms1} and this is illustrated in 
the string-like configuration of C3-1.  If we place atoms at the same 3/9 ML concentration to
a C3-1 sublattice, only one carbon atom is adsorbed into a C-1 site and the remaining
atoms form spontaneously a dimer.  This is demonstrated in configuration C3-2 of \fig{highdens2}.

At 4/9 ML coverage we have considered several different configurations:
C4-1 is very similar to CA-2 of \fig{geoms2} and the surface stress is again released by
expulsing two atoms towards vacuum while maintaining at least two other atoms on the C-1 site;
in C4-4, a complex reconstruction, where all carbon atoms are residing in 3-fold adsorption sites
occurs; at 6/9 ML concentration, even iron atoms are expelled from the surface in order to release
surface strain and to accommodate carbon atoms in the C-1 site; in C6-2, a C-Fe-C molecule is spontaneously formed.

\textbf{(111) surface.}~
In the case of the (111) we discussed how there is an 
abundance of open octahedral sites (C-1 and C-2 in \fig{geoms3}).  We also noted that
these sites are close to each other so that C$_2$ dimers are naturally formed.
Some of the most favorable coadsorption configurations at higher concentrations on the
(111) surface have been illustrated in \fig{highdens3}.  From the energetics in
\fig{highdens3}, we see that there is an energy cost when adsorbing more carbon on the surface,
but not as high as in the case of (110); no strong repulsion of carbon atoms takes place as
there are plenty of adsorption sites and the top layer atoms in (111) are flexible to move.
In (110), the C$_2$ dimers were formed above the topmost surface, resulting from expulsion of carbon atoms
from the C-1 sites, while in the (111), the carbon dimers are formed below the topmost layer,
in the very same sites where the atomic carbon is adsorbed.

\textbf{Surface energies.}~
The energetics of \tab{surfnrjs} and \figs{highdens}{highdens3}
are also shown in \fig{carboexes} as surface energies (employing \eq{surfes}).
This figure is not to be taken as an exact representation of surface
energies as function of carbon concentration; situations with ``graphenated'' surfaces
(see \refcite{Ref}{begtrup09}) and many possible carbon adsorption configurations are missing.
However, one can see some clear trends:
the steep rise in the surface energy as function of carbon concentration in the (110)
facet implies aggressive graphene formation on this facet;
the (100) facet forms stable carbidic phases near 0.5 ML = 0.08 1/\AA$^2$;
at high carbon concentrations the (111) facet gains in relative stability with respect to (100) facet.
This stabilization
can be understood in terms of frustration of the (100) facet, and from the bigger number
of adsorption sites available and the flexibility of the topmost iron atoms on the (111) facet,
as discussed above.

\subsection{Diffusion}

\begin{figure*}\centering
\putgraph{width=13cm}{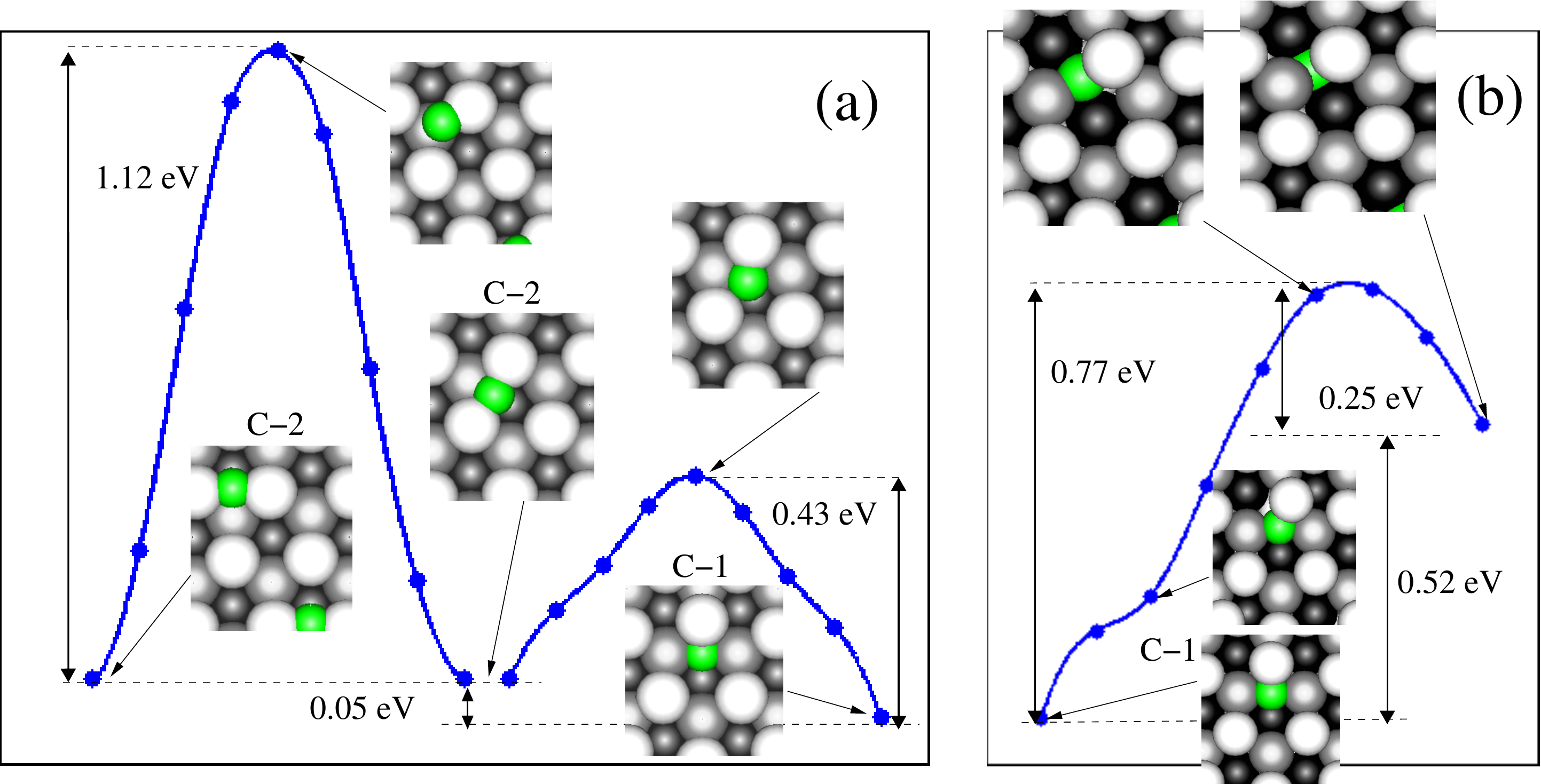}
\caption{
(a) Diffusion on (111) into the C-1 and C-2 sites (\fig{geoms2}) and (b) into a deeper sub-surface site.
\label{fig:nebs2}}
\end{figure*}

Diffusion on and into the (110) and (100) facets has been described earlier by
Jiang and Carter\cite{jiang05}.  We repeated their calculations for the activation barriers 
of carbon diffusion on the topmost iron layer and obtained identical results.
Here we report resuls for carbon diffusion on the (111) facet which, to our knowledge,
have not been calculated earlier using \abit methods.

When studying sub-surface adsorption and diffusion, one must
keep in mind the deep interlayer relaxations of pure iron slabs and the elongation/contraction 
of distances (A) and (B) (\fig{bcciron}) which might propagate far in the lattice.
Then a thorough investigation of activation energies and energy cost for carbon atom to enter 
the bulk would require calculations with very thick slabs, considering increasing
adsorption depths until bulk values are recovered.

In \refcite{Ref}{jiang05} barriers and energetics for (110) and (100) were calculated down 
to the first sub-surface layer.  As discussed in \sct{atomic} for (111) facet, the 111/C-1 site 
of \fig{geoms3} can be classified as
``semi'' sub-surface site.  In our calculations, we have considered also the first ``true'' sub-surface
site (i.e. a site that has bulk-like coordination), but have not pursued
the calculation of the diffusion barrier when going very deep inside the slab as the limits of our finite
slab are quickly reached.

Carbon diffusion between neighboring C-2 sites and between the C-1 and C-2 sites has been illustrated in \figa{nebs2}{a}.
Diffusion from C-2 to deeper inside the slab and reaching a site where carbon has similar coordination
as in the bulk octahedral site, is illustrated in \figa{nebs2}{b}.  This site, not reported
in \fig{geoms3}, should not be confused with the C-4 adsorbate which lies even deeper inside the slab.

The activation energy E$_a$=1.12 eV for surface diffusion reported here for the (111) surface is smaller than for the (100) surface (1.45 eV), but slightly bigger than for the (110) surface (0.96 eV)\simplecite{jiang05}.
Going from C-2 site to the ``semi'' sub-surface site C-1 has a very low barrier of 0.43 eV.
Going deeper inside the slab has a barrier of 0.77 eV. 
As illustrated in the insets of \figa{nebs2}{b}, in the transition state of the minimum
energy path, one atom in the topmost iron layer is pushed away from the carbon atom. 
However, the barrier is low (0.77 eV) because the iron atoms in the (111) topmost layer are low coordinated and
flexible to move.
We identify the 0.77 eV barrier tentatively as the activation energy for carbon atom to enter the surface
and it is smaller than those reported for (110) (1.18 eV) and (100) (1.47 eV).
Moreover, from \tab{surfnrjs} the energy difference between the  111/C-1 and 111/C-4 sites,
is only 0.33 eV, which is smaller than the same energy difference in
(110) (0.62 eV) and (100) (1.19 eV)\simplecite{jiang05}.
There are then clear indications that the exchange of carbon atoms between the topmost and
deeper layers is easier on the (111) facet than on (110) and (100) facets.

\section{Discussions and Conclusions}
\label{sec:conc}
This work was motivated by the recent \insi studies of carbon nanotube growth
from ``large'' iron nanoparticles where different nanoparticle facets
seem to behave in a very different manner.  The facets studied in this
work correspond to those identified in \refcite{Ref}{begtrup09}, namely the
\airon~ (110), (100) and (111) surfaces.  We have studied the effect of adsorbing
increasing amounts of carbon on these surfaces 

The repulsion among adsorbed carbon atoms on the (110) facet is strong and because of this,
carbon atoms are expulsed from the optimal adsorption sites when carbon concentration is
increased. This may happen already at the relatively low 
0.22 ML concentration, resulting in dimer and graphene formation.

The (100) facet behaves in a very different way, as stable, carbon rich structures are formed
near 0.5 ML concentration.   When approaching 1 ML carbon concentration, (100)
is destabilized due to frustration of the top layer iron atoms, which can be
released by dimer and graphitic material formation.

The (111) surface behaves again in a distinct manner; the surface energy is rather
insensitive to the amount of carbon adsorbed on it (at similar atoms/$\AA^2$
concentrations where (100) becomes unstable).  This results from the abundance
of adsorption sites and from the flexibility of the topmost iron atom layer.

Carbon nanotube and graphene formation are known to be highly kinetic processes, so one must be cautious
when relating the present work - based mainly on total energies of different adsorbed carbon 
concentrations - to these processes.  However, the arguments concerning the nature of adsorbate
repulsions and surface iron atom frustration on the different facets are valid also in a 
dynamical situation and at higher temperatures, as long as the nanoparticles are crystalline.

The sudden carbon ``supersaturation'' at the (100) facet near 1 ML carbon concentration might be related
to the lift of graphitic caps from this facet
as more carbon is injected into the nanoparticle
as seen in the \insi experiment\cite{begtrup09}.
When carbon feedstock is exhausted in the experiment,
the (111) facets start to shrink, eventually disappearing\cite{begtrup09}.
This is very likely related to our computational results: the relative stabilities of
(100) and (111) were observed to be sensitive to the amount of adsorbed carbon. At low carbon concentration
(100) gains in stability while at high concentrations, (111) becomes equally stable.
The exchange of carbon between the topmost iron layer and the subsurface layers was seen to be easier
on (111) than on the other facets.  Carbon dimer formation was observed to be most favorable on (111).
This could be related to the typical TEM image of a CVD grown multi-walled carbon nanotubes
that show several graphitic layers emerging from the (111) facet\cite{begtrup09}.

These last observations can be important when trying to understand the growth mechanisms
of carbon nanotubes in general:
in a pure \airon~ nanoparticle, the portion of (111) facets is very small \cite{blonski07}.  As carbon concentration on the nanoparticle surface gets higher, (111) facets will be stabilized.  Once these facets have been established, graphitic material growth from them can proceed as they favor C$_2$ formation and as the movement of carbon atoms
between sub-surface and surface is easier.
However, in order to make this idea more solid, more investigation about the diffusion of carbon and kinetics of graphene growth on the (111) facet must be performed.

To summarize, carbon concentrations of up to one monolayer were studied on \airon~facets.  Such aspects as
repulsion between carbon adsorbates on the (110) the frustration of iron atoms on the (100) surface
and the dimer formation on (111) facet together with their effect on the surface energies were discussed.  Diffusion on and into the (111) facet was studied.  Our findings were related to a recent \insi~study where the appearance of (111) facets correlates with increased carbon concentration.  A general idea where increased carbon concentration stabilizes the (111) facets followed by growth of graphitic material from these facets was proposed.

\section{Acknowledgements}
We wish to thank the Center for Scientific Computing, for use of its computational resources.
This work has been supported in part by the European Commission under 
the Framework Programme (STREP project BNC Tubes, contract number NMP4-CT-2006-03350)
and the Academy of Finland through its Centre of Excellence Programme (2006-2011).

\bibliographystyle{apsrev-oma} 

\begin{thebibliography}{38}
\expandafter\ifx\csname natexlab\endcsname\relax\def\natexlab#1{#1}\fi
\expandafter\ifx\csname bibnamefont\endcsname\relax
  \def\bibnamefont#1{#1}\fi
\expandafter\ifx\csname bibfnamefont\endcsname\relax
  \def\bibfnamefont#1{#1}\fi
\expandafter\ifx\csname citenamefont\endcsname\relax
  \def\citenamefont#1{#1}\fi
\expandafter\ifx\csname url\endcsname\relax
  \def\url#1{\texttt{#1}}\fi
\expandafter\ifx\csname urlprefix\endcsname\relax\def\urlprefix{URL }\fi
\providecommand{\bibinfo}[2]{#2}
\providecommand{\eprint}[2][]{\url{#2}}

\bibitem[{\citenamefont{Sharma and Iqbal}(2004)}]{sharma04}
\bibinfo{author}{\bibfnamefont{R.}~\bibnamefont{Sharma}} \bibnamefont{and}
  \bibinfo{author}{\bibfnamefont{Z.}~\bibnamefont{Iqbal}},
  \bibinfo{journal}{Appl. Phys. Lett.} \textbf{\bibinfo{volume}{84}},
  \bibinfo{pages}{990} (\bibinfo{year}{2004}).

\bibitem[{\citenamefont{Helveg et~al.}(2004)\citenamefont{Helveg, Lopez-Cartes,
  Sehested, Hansen, Clausen, Rostrup-Nielsen, Abild-Pedersen, and
  N\o{}rskov}}]{helveg04}
\bibinfo{author}{\bibfnamefont{S.}~\bibnamefont{Helveg}},
  \bibinfo{author}{\bibfnamefont{C.}~\bibnamefont{Lopez-Cartes}},
  \bibinfo{author}{\bibfnamefont{J.}~\bibnamefont{Sehested}},
  \bibinfo{author}{\bibfnamefont{P.~L.} \bibnamefont{Hansen}},
  \bibinfo{author}{\bibfnamefont{B.~S.} \bibnamefont{Clausen}},
  \bibinfo{author}{\bibfnamefont{J.~R.} \bibnamefont{Rostrup-Nielsen}},
  \bibinfo{author}{\bibfnamefont{F.}~\bibnamefont{Abild-Pedersen}},
  \bibnamefont{and} \bibinfo{author}{\bibfnamefont{J.~K.}
  \bibnamefont{N\o{}rskov}}, \bibinfo{journal}{Nature}
  \textbf{\bibinfo{volume}{427}}, \bibinfo{pages}{426} (\bibinfo{year}{2004}).

\bibitem[{\citenamefont{Hofmann et~al.}(2007)\citenamefont{Hofmann, Sharma,
  Ducati, Du, Mattevi, Cepek, Cantoro, Pisana, Parvez, Cervantes-Sodi
  et~al.}}]{hofmann07}
\bibinfo{author}{\bibfnamefont{S.}~\bibnamefont{Hofmann}},
  \bibinfo{author}{\bibfnamefont{R.}~\bibnamefont{Sharma}},
  \bibinfo{author}{\bibfnamefont{C.}~\bibnamefont{Ducati}},
  \bibinfo{author}{\bibfnamefont{G.}~\bibnamefont{Du}},
  \bibinfo{author}{\bibfnamefont{C.}~\bibnamefont{Mattevi}},
  \bibinfo{author}{\bibfnamefont{C.}~\bibnamefont{Cepek}},
  \bibinfo{author}{\bibfnamefont{M.}~\bibnamefont{Cantoro}},
  \bibinfo{author}{\bibfnamefont{S.}~\bibnamefont{Pisana}},
  \bibinfo{author}{\bibfnamefont{A.}~\bibnamefont{Parvez}},
  \bibinfo{author}{\bibfnamefont{F.}~\bibnamefont{Cervantes-Sodi}},
  \bibnamefont{et~al.}, \bibinfo{journal}{Nano Lett.}
  \textbf{\bibinfo{volume}{7}}, \bibinfo{pages}{602} (\bibinfo{year}{2007}).

\bibitem[{\citenamefont{Jensen et~al.}(2005)\citenamefont{Jensen, Mickelson,
  Han, and Zettl}}]{jensen05}
\bibinfo{author}{\bibfnamefont{K.}~\bibnamefont{Jensen}},
  \bibinfo{author}{\bibfnamefont{W.}~\bibnamefont{Mickelson}},
  \bibinfo{author}{\bibfnamefont{W.}~\bibnamefont{Han}}, \bibnamefont{and}
  \bibinfo{author}{\bibfnamefont{A.}~\bibnamefont{Zettl}},
  \bibinfo{journal}{Appl. Phys. Lett.} \textbf{\bibinfo{volume}{86}},
  \bibinfo{eid}{173107} (pages~\bibinfo{numpages}{3}) (\bibinfo{year}{2005}).

\bibitem[{\citenamefont{Abild-Pedersen
  et~al.}(2006)\citenamefont{Abild-Pedersen, N\o{}rskov, Rostrup-Nielsen,
  Sehested, and Helveg}}]{abildpedersen06}
\bibinfo{author}{\bibfnamefont{F.}~\bibnamefont{Abild-Pedersen}},
  \bibinfo{author}{\bibfnamefont{J.~K.} \bibnamefont{N\o{}rskov}},
  \bibinfo{author}{\bibfnamefont{J.~R.} \bibnamefont{Rostrup-Nielsen}},
  \bibinfo{author}{\bibfnamefont{J.}~\bibnamefont{Sehested}}, \bibnamefont{and}
  \bibinfo{author}{\bibfnamefont{S.}~\bibnamefont{Helveg}},
  \bibinfo{journal}{Phys. Rev. B} \textbf{\bibinfo{volume}{73}},
  \bibinfo{eid}{115419} (\bibinfo{year}{2006}).

\bibitem[{\citenamefont{Begtrup et~al.}(2009)\citenamefont{Begtrup, Gannett,
  Meyer, Yuzvinsky, Ertekin, Grossman, and Zettl}}]{begtrup09}
\bibinfo{author}{\bibfnamefont{G.~E.} \bibnamefont{Begtrup}},
  \bibinfo{author}{\bibfnamefont{W.}~\bibnamefont{Gannett}},
  \bibinfo{author}{\bibfnamefont{J.~C.} \bibnamefont{Meyer}},
  \bibinfo{author}{\bibfnamefont{T.~D.} \bibnamefont{Yuzvinsky}},
  \bibinfo{author}{\bibfnamefont{E.}~\bibnamefont{Ertekin}},
  \bibinfo{author}{\bibfnamefont{J.~C.} \bibnamefont{Grossman}},
  \bibnamefont{and} \bibinfo{author}{\bibfnamefont{A.}~\bibnamefont{Zettl}},
  \bibinfo{journal}{Phys. Rev. B} \textbf{\bibinfo{volume}{79}},
  \bibinfo{eid}{205409} (pages~\bibinfo{numpages}{6}) (\bibinfo{year}{2009}).

\bibitem[{\citenamefont{Rodriguez-Manzo
  et~al.}(2007)\citenamefont{Rodriguez-Manzo, Terrones, Terrones, Kroto, Sun,
  and Banhart}}]{rodriguezmanzo07}
\bibinfo{author}{\bibfnamefont{J.~A.} \bibnamefont{Rodriguez-Manzo}},
  \bibinfo{author}{\bibfnamefont{M.}~\bibnamefont{Terrones}},
  \bibinfo{author}{\bibfnamefont{H.}~\bibnamefont{Terrones}},
  \bibinfo{author}{\bibfnamefont{H.~W.} \bibnamefont{Kroto}},
  \bibinfo{author}{\bibfnamefont{L.}~\bibnamefont{Sun}}, \bibnamefont{and}
  \bibinfo{author}{\bibfnamefont{F.}~\bibnamefont{Banhart}},
  \bibinfo{journal}{Nature Nanotechnology} \textbf{\bibinfo{volume}{2}},
  \bibinfo{pages}{307} (\bibinfo{year}{2007}).

\bibitem[{\citenamefont{Yoshida et~al.}(2008)\citenamefont{Yoshida, Takeda,
  Uchiyama, Kohno, and Homma}}]{yoshida08}
\bibinfo{author}{\bibfnamefont{H.}~\bibnamefont{Yoshida}},
  \bibinfo{author}{\bibfnamefont{S.}~\bibnamefont{Takeda}},
  \bibinfo{author}{\bibfnamefont{T.}~\bibnamefont{Uchiyama}},
  \bibinfo{author}{\bibfnamefont{H.}~\bibnamefont{Kohno}}, \bibnamefont{and}
  \bibinfo{author}{\bibfnamefont{Y.}~\bibnamefont{Homma}},
  \bibinfo{journal}{Nano Lett.} \textbf{\bibinfo{volume}{8}},
  \bibinfo{pages}{2082} (\bibinfo{year}{2008}).

\bibitem[{\citenamefont{Rodr{\'i}guez-Manzo
  et~al.}(2009)\citenamefont{Rodr{\'i}guez-Manzo, Janowska, Pham-Huu, Tolvanen,
  Krasheninnikov, Nordlund, and Banhart}}]{rodriguezmanzo09}
\bibinfo{author}{\bibfnamefont{J.~A.} \bibnamefont{Rodr{\'i}guez-Manzo}},
  \bibinfo{author}{\bibfnamefont{I.}~\bibnamefont{Janowska}},
  \bibinfo{author}{\bibfnamefont{C.}~\bibnamefont{Pham-Huu}},
  \bibinfo{author}{\bibfnamefont{A.}~\bibnamefont{Tolvanen}},
  \bibinfo{author}{\bibfnamefont{A.~V.} \bibnamefont{Krasheninnikov}},
  \bibinfo{author}{\bibfnamefont{K.}~\bibnamefont{Nordlund}}, \bibnamefont{and}
  \bibinfo{author}{\bibfnamefont{F.}~\bibnamefont{Banhart}},
  \bibinfo{journal}{Small} \textbf{\bibinfo{volume}{5}}, \bibinfo{pages}{2710}
  (\bibinfo{year}{2009}).

\bibitem[{\citenamefont{Sharma et~al.}(2009)\citenamefont{Sharma, Moore, Rez,
  and Treacy}}]{sharma09}
\bibinfo{author}{\bibfnamefont{R.}~\bibnamefont{Sharma}},
  \bibinfo{author}{\bibfnamefont{E.}~\bibnamefont{Moore}},
  \bibinfo{author}{\bibfnamefont{P.}~\bibnamefont{Rez}}, \bibnamefont{and}
  \bibinfo{author}{\bibfnamefont{M.~M.~J.} \bibnamefont{Treacy}},
  \bibinfo{journal}{Nano Lett.} \textbf{\bibinfo{volume}{9}},
  \bibinfo{pages}{689} (\bibinfo{year}{2009}).

\bibitem[{\citenamefont{Bengaard et~al.}(2002)\citenamefont{Bengaard,
  N\o{}rskov, Sehested, Clausen, Nielsen, Molenbroek, and
  Rostrup-Nielsen}}]{bengaard02}
\bibinfo{author}{\bibfnamefont{H.~S.} \bibnamefont{Bengaard}},
  \bibinfo{author}{\bibfnamefont{J.~K.} \bibnamefont{N\o{}rskov}},
  \bibinfo{author}{\bibfnamefont{J.}~\bibnamefont{Sehested}},
  \bibinfo{author}{\bibfnamefont{B.~S.} \bibnamefont{Clausen}},
  \bibinfo{author}{\bibfnamefont{L.~P.} \bibnamefont{Nielsen}},
  \bibinfo{author}{\bibfnamefont{A.~M.} \bibnamefont{Molenbroek}},
  \bibnamefont{and} \bibinfo{author}{\bibfnamefont{J.~R.}
  \bibnamefont{Rostrup-Nielsen}}, \bibinfo{journal}{J. of Catalysis}
  \textbf{\bibinfo{volume}{209}}, \bibinfo{pages}{365} (\bibinfo{year}{2002}).

\bibitem[{\citenamefont{Harutyunyan et~al.}(2008)\citenamefont{Harutyunyan,
  Awasthi, Jiang, Setyawan, Mora, Tokune, Bolton, and
  Curtarolo}}]{harutyunyan08}
\bibinfo{author}{\bibfnamefont{A.~R.} \bibnamefont{Harutyunyan}},
  \bibinfo{author}{\bibfnamefont{N.}~\bibnamefont{Awasthi}},
  \bibinfo{author}{\bibfnamefont{A.}~\bibnamefont{Jiang}},
  \bibinfo{author}{\bibfnamefont{W.}~\bibnamefont{Setyawan}},
  \bibinfo{author}{\bibfnamefont{E.}~\bibnamefont{Mora}},
  \bibinfo{author}{\bibfnamefont{T.}~\bibnamefont{Tokune}},
  \bibinfo{author}{\bibfnamefont{K.}~\bibnamefont{Bolton}}, \bibnamefont{and}
  \bibinfo{author}{\bibfnamefont{S.}~\bibnamefont{Curtarolo}},
  \bibinfo{journal}{Phys. Rev. Lett.} \textbf{\bibinfo{volume}{100}},
  \bibinfo{eid}{195502} (pages~\bibinfo{numpages}{4}) (\bibinfo{year}{2008}).

\bibitem[{\citenamefont{Barnard and Zapol}(2004{\natexlab{a}})}]{barnard04}
\bibinfo{author}{\bibfnamefont{A.~S.} \bibnamefont{Barnard}} \bibnamefont{and}
  \bibinfo{author}{\bibfnamefont{P.}~\bibnamefont{Zapol}},
  \bibinfo{journal}{The J. of Chem. Phys.} \textbf{\bibinfo{volume}{121}},
  \bibinfo{pages}{4276} (\bibinfo{year}{2004}{\natexlab{a}}).

\bibitem[{\citenamefont{Hofmann et~al.}(2005)\citenamefont{Hofmann, Csanyi,
  Ferrari, Payne, and Robertson}}]{hofmann05}
\bibinfo{author}{\bibfnamefont{S.}~\bibnamefont{Hofmann}},
  \bibinfo{author}{\bibfnamefont{G.}~\bibnamefont{Csanyi}},
  \bibinfo{author}{\bibfnamefont{A.~C.} \bibnamefont{Ferrari}},
  \bibinfo{author}{\bibfnamefont{M.~C.} \bibnamefont{Payne}}, \bibnamefont{and}
  \bibinfo{author}{\bibfnamefont{J.}~\bibnamefont{Robertson}},
  \bibinfo{journal}{Phys. Rev. Lett.} \textbf{\bibinfo{volume}{95}},
  \bibinfo{eid}{036101} (pages~\bibinfo{numpages}{4}) (\bibinfo{year}{2005}).

\bibitem[{\citenamefont{B\l\'{o}nski and Kiejna}(2004)}]{blonski03}
\bibinfo{author}{\bibfnamefont{P.}~\bibnamefont{B\l\'{o}nski}}
  \bibnamefont{and} \bibinfo{author}{\bibfnamefont{A.}~\bibnamefont{Kiejna}},
  \bibinfo{journal}{Vacuum} \textbf{\bibinfo{volume}{74}}, \bibinfo{pages}{179}
  (\bibinfo{year}{2004}).

\bibitem[{\citenamefont{B\l\'{o}nski and Kiejna}(2007)}]{blonski07}
\bibinfo{author}{\bibfnamefont{P.}~\bibnamefont{B\l\'{o}nski}}
  \bibnamefont{and} \bibinfo{author}{\bibfnamefont{A.}~\bibnamefont{Kiejna}},
  \bibinfo{journal}{Surface Science} \textbf{\bibinfo{volume}{601}},
  \bibinfo{pages}{123 } (\bibinfo{year}{2007}).

\bibitem[{\citenamefont{Jiang and Carter}(2005)}]{jiang05}
\bibinfo{author}{\bibfnamefont{D.~E.} \bibnamefont{Jiang}} \bibnamefont{and}
  \bibinfo{author}{\bibfnamefont{E.~A.} \bibnamefont{Carter}},
  \bibinfo{journal}{Phys. Rev. B} \textbf{\bibinfo{volume}{71}},
  \bibinfo{pages}{045402} (\bibinfo{year}{2005}).

\bibitem[{\citenamefont{Blum et~al.}(2003)\citenamefont{Blum, Schmidt, Meier,
  Hammer, and Heinz}}]{blum03}
\bibinfo{author}{\bibfnamefont{V.}~\bibnamefont{Blum}},
  \bibinfo{author}{\bibfnamefont{A.}~\bibnamefont{Schmidt}},
  \bibinfo{author}{\bibfnamefont{W.}~\bibnamefont{Meier}},
  \bibinfo{author}{\bibfnamefont{L.}~\bibnamefont{Hammer}}, \bibnamefont{and}
  \bibinfo{author}{\bibfnamefont{K.}~\bibnamefont{Heinz}}, \bibinfo{journal}{J.
  of Physics: Condensed Matter} \textbf{\bibinfo{volume}{15}},
  \bibinfo{pages}{3517} (\bibinfo{year}{2003}).

\bibitem[{\citenamefont{Panaccione et~al.}(2006)\citenamefont{Panaccione,
  Fujii, Vobornik, Trimarchi, Binggeli, Goldoni, Larciprete, and
  Rossi}}]{panaccione06}
\bibinfo{author}{\bibfnamefont{G.}~\bibnamefont{Panaccione}},
  \bibinfo{author}{\bibfnamefont{J.}~\bibnamefont{Fujii}},
  \bibinfo{author}{\bibfnamefont{I.}~\bibnamefont{Vobornik}},
  \bibinfo{author}{\bibfnamefont{G.}~\bibnamefont{Trimarchi}},
  \bibinfo{author}{\bibfnamefont{N.}~\bibnamefont{Binggeli}},
  \bibinfo{author}{\bibfnamefont{A.}~\bibnamefont{Goldoni}},
  \bibinfo{author}{\bibfnamefont{R.}~\bibnamefont{Larciprete}},
  \bibnamefont{and} \bibinfo{author}{\bibfnamefont{G.}~\bibnamefont{Rossi}},
  \bibinfo{journal}{Phys. Rev. B} \textbf{\bibinfo{volume}{73}},
  \bibinfo{pages}{035431} (\bibinfo{year}{2006}).

\bibitem[{\citenamefont{Tan et~al.}(2010)\citenamefont{Tan, Zhou, Liu, Peng,
  and Zhao}}]{tan10}
\bibinfo{author}{\bibfnamefont{X.}~\bibnamefont{Tan}},
  \bibinfo{author}{\bibfnamefont{J.}~\bibnamefont{Zhou}},
  \bibinfo{author}{\bibfnamefont{F.}~\bibnamefont{Liu}},
  \bibinfo{author}{\bibfnamefont{Y.}~\bibnamefont{Peng}}, \bibnamefont{and}
  \bibinfo{author}{\bibfnamefont{B.}~\bibnamefont{Zhao}}, \bibinfo{journal}{The
  European Physical Journal B} \textbf{\bibinfo{volume}{74}},
  \bibinfo{pages}{555} (\bibinfo{year}{2010}).

\bibitem[{\citenamefont{Chiou and Carter}(2003)}]{chiou03}
\bibinfo{author}{\bibfnamefont{W.~C.} \bibnamefont{Chiou}} \bibnamefont{and}
  \bibinfo{author}{\bibfnamefont{E.~A.} \bibnamefont{Carter}},
  \bibinfo{journal}{Surface Science} \textbf{\bibinfo{volume}{530}},
  \bibinfo{pages}{88 } (\bibinfo{year}{2003}).

\bibitem[{\citenamefont{Tilley}(2004)}]{tilley}
\bibinfo{author}{\bibfnamefont{R.~J.~D.} \bibnamefont{Tilley}},
  \emph{\bibinfo{title}{Understanding Solids: The Science of Materials}}
  (\bibinfo{publisher}{Wiley}, \bibinfo{year}{2004}).

\bibitem[{\citenamefont{Rollmann et~al.}(2007)\citenamefont{Rollmann, Gruner,
  Hucht, Meyer, Entel, Tiago, and Chelikowsky}}]{rollman07}
\bibinfo{author}{\bibfnamefont{G.}~\bibnamefont{Rollmann}},
  \bibinfo{author}{\bibfnamefont{M.~E.} \bibnamefont{Gruner}},
  \bibinfo{author}{\bibfnamefont{A.}~\bibnamefont{Hucht}},
  \bibinfo{author}{\bibfnamefont{R.}~\bibnamefont{Meyer}},
  \bibinfo{author}{\bibfnamefont{P.}~\bibnamefont{Entel}},
  \bibinfo{author}{\bibfnamefont{M.~L.} \bibnamefont{Tiago}}, \bibnamefont{and}
  \bibinfo{author}{\bibfnamefont{J.~R.} \bibnamefont{Chelikowsky}},
  \bibinfo{journal}{Phys. Rev. Lett.} \textbf{\bibinfo{volume}{99}},
  \bibinfo{pages}{083402} (\bibinfo{year}{2007}).

\bibitem[{\citenamefont{Postnikov et~al.}(2003)\citenamefont{Postnikov, Entel,
  and Soler}}]{postnikov03}
\bibinfo{author}{\bibfnamefont{A.}~\bibnamefont{Postnikov}},
  \bibinfo{author}{\bibfnamefont{P.}~\bibnamefont{Entel}}, \bibnamefont{and}
  \bibinfo{author}{\bibfnamefont{J.}~\bibnamefont{Soler}},
  \bibinfo{journal}{European Physical Journal D} \textbf{\bibinfo{volume}{25}},
  \bibinfo{pages}{261} (\bibinfo{year}{2003}).

\bibitem[{\citenamefont{Perdew et~al.}(1996)\citenamefont{Perdew, Burke, and
  Ernzerhof}}]{gga}
\bibinfo{author}{\bibfnamefont{J.~P.} \bibnamefont{Perdew}},
  \bibinfo{author}{\bibfnamefont{K.}~\bibnamefont{Burke}}, \bibnamefont{and}
  \bibinfo{author}{\bibfnamefont{M.}~\bibnamefont{Ernzerhof}},
  \bibinfo{journal}{Phys. Rev. Lett.} \textbf{\bibinfo{volume}{77}},
  \bibinfo{pages}{3865} (\bibinfo{year}{1996}).

\bibitem[{\citenamefont{Jiang and Carter}(2003)}]{jiang03}
\bibinfo{author}{\bibfnamefont{D.~E.} \bibnamefont{Jiang}} \bibnamefont{and}
  \bibinfo{author}{\bibfnamefont{E.~A.} \bibnamefont{Carter}},
  \bibinfo{journal}{Phys. Rev. B} \textbf{\bibinfo{volume}{67}},
  \bibinfo{pages}{214103} (\bibinfo{year}{2003}).

\bibitem[{\citenamefont{Barnard and Zapol}(2004{\natexlab{b}})}]{barnard04_2}
\bibinfo{author}{\bibfnamefont{A.~S.} \bibnamefont{Barnard}} \bibnamefont{and}
  \bibinfo{author}{\bibfnamefont{P.}~\bibnamefont{Zapol}},
  \bibinfo{journal}{Phys. Rev. B} \textbf{\bibinfo{volume}{70}},
  \bibinfo{pages}{235403} (\bibinfo{year}{2004}{\natexlab{b}}).

\bibitem[{\citenamefont{Boettger}(1996)}]{boettger96}
\bibinfo{author}{\bibfnamefont{J.~C.} \bibnamefont{Boettger}},
  \bibinfo{journal}{Phys. Rev. B} \textbf{\bibinfo{volume}{53}},
  \bibinfo{pages}{13133} (\bibinfo{year}{1996}).

\bibitem[{\citenamefont{Kresse and Furthm\"uller}(1996)}]{vasp2}
\bibinfo{author}{\bibfnamefont{G.}~\bibnamefont{Kresse}} \bibnamefont{and}
  \bibinfo{author}{\bibfnamefont{J.}~\bibnamefont{Furthm\"uller}},
  \bibinfo{journal}{Phys. Rev. B} \textbf{\bibinfo{volume}{54}},
  \bibinfo{pages}{11169} (\bibinfo{year}{1996}).

\bibitem[{\citenamefont{Kresse and Joubert}(1999)}]{vasp3}
\bibinfo{author}{\bibfnamefont{G.}~\bibnamefont{Kresse}} \bibnamefont{and}
  \bibinfo{author}{\bibfnamefont{D.}~\bibnamefont{Joubert}},
  \bibinfo{journal}{Phys. Rev. B} \textbf{\bibinfo{volume}{59}},
  \bibinfo{pages}{1758} (\bibinfo{year}{1999}).

\bibitem[{\citenamefont{Bl\"ochl}(1994)}]{paw}
\bibinfo{author}{\bibfnamefont{P.~E.} \bibnamefont{Bl\"ochl}},
  \bibinfo{journal}{Phys. Rev. B} \textbf{\bibinfo{volume}{50}},
  \bibinfo{pages}{17953} (\bibinfo{year}{1994}).

\bibitem[{\citenamefont{Monkhorst and Pack}(1976)}]{mp}
\bibinfo{author}{\bibfnamefont{H.~J.} \bibnamefont{Monkhorst}}
  \bibnamefont{and} \bibinfo{author}{\bibfnamefont{J.~D.} \bibnamefont{Pack}},
  \bibinfo{journal}{Phys. Rev. B} \textbf{\bibinfo{volume}{13}},
  \bibinfo{pages}{5188} (\bibinfo{year}{1976}).

\bibitem[{\citenamefont{Riikonen et~al.}(2009)\citenamefont{Riikonen, Foster,
  Krasheninnikov, and Nieminen}}]{riikonen09}
\bibinfo{author}{\bibfnamefont{S.}~\bibnamefont{Riikonen}},
  \bibinfo{author}{\bibfnamefont{A.~S.} \bibnamefont{Foster}},
  \bibinfo{author}{\bibfnamefont{A.~V.} \bibnamefont{Krasheninnikov}},
  \bibnamefont{and} \bibinfo{author}{\bibfnamefont{R.~M.}
  \bibnamefont{Nieminen}}, \bibinfo{journal}{Phys. Rev. B}
  \textbf{\bibinfo{volume}{80}}, \bibinfo{eid}{155429} (\bibinfo{year}{2009}).

\bibitem[{\citenamefont{Methfessel and Paxton}(1989)}]{mpax}
\bibinfo{author}{\bibfnamefont{M.}~\bibnamefont{Methfessel}} \bibnamefont{and}
  \bibinfo{author}{\bibfnamefont{A.~T.} \bibnamefont{Paxton}},
  \bibinfo{journal}{Phys. Rev. B} \textbf{\bibinfo{volume}{40}},
  \bibinfo{pages}{3616} (\bibinfo{year}{1989}).

\bibitem[{\citenamefont{Wirtz and Rubio}(2004)}]{wirtz04}
\bibinfo{author}{\bibfnamefont{L.}~\bibnamefont{Wirtz}} \bibnamefont{and}
  \bibinfo{author}{\bibfnamefont{A.}~\bibnamefont{Rubio}},
  \bibinfo{journal}{Solid State Communications} \textbf{\bibinfo{volume}{131}},
  \bibinfo{pages}{141 } (\bibinfo{year}{2004}).

\bibitem[{\citenamefont{Gui et~al.}(2008)\citenamefont{Gui, Li, and
  Zhong}}]{gui08}
\bibinfo{author}{\bibfnamefont{G.}~\bibnamefont{Gui}},
  \bibinfo{author}{\bibfnamefont{J.}~\bibnamefont{Li}}, \bibnamefont{and}
  \bibinfo{author}{\bibfnamefont{J.}~\bibnamefont{Zhong}},
  \bibinfo{journal}{Phys. Rev. B} \textbf{\bibinfo{volume}{78}},
  \bibinfo{pages}{075435} (\bibinfo{year}{2008}).

\bibitem[{\citenamefont{Henkelman et~al.}(2000)\citenamefont{Henkelman,
  Uberuaga, and J\'{o}nsson}}]{henkelman00}
\bibinfo{author}{\bibfnamefont{G.}~\bibnamefont{Henkelman}},
  \bibinfo{author}{\bibfnamefont{B.~P.} \bibnamefont{Uberuaga}},
  \bibnamefont{and}
  \bibinfo{author}{\bibfnamefont{H.}~\bibnamefont{J\'{o}nsson}},
  \bibinfo{journal}{The J. of Chem. Phys.} \textbf{\bibinfo{volume}{113}},
  \bibinfo{pages}{9901} (\bibinfo{year}{2000}).

\bibitem[{\citenamefont{Kittel}(1986)}]{kittel}
\bibinfo{author}{\bibfnamefont{C.}~\bibnamefont{Kittel}},
  \emph{\bibinfo{title}{Introduction to Solid State Physics, 6.th edition.}}
  (\bibinfo{publisher}{Wiley, New York}, \bibinfo{year}{1986}).

\end{thebibliography}

\end{document}